\documentclass[11pt]{article}
\usepackage{graphicx}
\graphicspath{{./}{Figures/}}
\usepackage[section]{placeins}

\usepackage{booktabs}
\usepackage{tabularx}
\usepackage{array}
\usepackage{amsmath}
\usepackage[T1]{fontenc}
\usepackage[margin=1in]{geometry}
\usepackage[hidelinks]{hyperref}
\usepackage{xcolor}
\usepackage{amssymb}
\newcommand{\yes}{\textcolor{green!50!black}{\checkmark}}
\usepackage{authblk} \author[1]{Aakash Shailesh Oza} \author[1]{Ryan F. Johnson} \author[1]{Amir Barati Farimani\thanks{Corresponding author: \href{mailto:barati@cmu.edu}{barati@cmu.edu}.}} \affil[1]{Department of Mechanical Engineering, Carnegie Mellon University, 5000 Forbes Avenue, Pittsburgh, PA 15213, USA} 

\title{CFD Forge: A Physics-Constrained Agentic Framework for Autonomous CFD}

\date{}

\begin{document}

\maketitle

\begin{abstract}
Setting up, running, and assessing a computational fluid dynamics (CFD) simulation requires substantial manual effort, from meshing and solver setup to diagnosis and interpretation, and repeating it across design variants slows engineering iteration. We present CFD Forge, an agentic framework built around OpenFOAM that carries an engineering request through case construction, meshing, solver execution, diagnostics, and reporting. A large language model interprets the request, reads the simulation evidence, identifies likely causes of problems, and proposes actions from a closed set. Automation is only useful if its results can be trusted, so acceptance is not left to the model: deterministic checks, registered for each type of flow, measure explicit criteria and decide. We demonstrate the framework on a converging--diverging nozzle and a two-dimensional forward-facing step, including geometry and operating changes requested in text. A turbulent cube, run outside the agent loop, shows why the checks matter: the solver completed and the mean drag settled, but the lateral-force oscillations were still growing, so a retrospective assessment withheld acceptance. To test this separation, we replayed 36 recorded simulation states, some with deliberately altered diagnostics, through the same model in two setups. Judging from the evidence without the check results, the model accepted 12 of the 13 detectable faults in all three repeats; with CFD Forge's checks deciding, none was accepted. The two setups also differ in prompting, and a fixed rule reached the same verdicts, so the model's value lies in the open-ended work rather than the verdict. Case files, recorded model calls, and decisions are retained for audit. A model that does the work, paired with checks that decide what is accepted, is a step toward automated CFD that can be trusted in high-throughput design and active-learning loops.
\end{abstract}

\section{Introduction}

Computational fluid dynamics (CFD) is central to engineering analysis, but obtaining a useful result requires a sequence of decisions and operations. An analyst prepares a CAD model or parameterized geometry, defines the domain, generates a mesh, selects physical models, and assigns boundary and initial conditions. Solver execution is then followed by diagnostics and interpretation, with the final aim of reporting results that range from scientific discovery to engineering decisions. Although individual steps can be scripted, coordinating them requires extensive experience, so much so that the need for workflow automation was identified in the NASA CFD Vision 2030 study to support routine computational analysis and design~\cite{slotnick2014vision}.

The effort required for CFD analysis creates a bottleneck in high-throughput screening and iterative design, where many geometries or operating conditions must be evaluated consistently \cite{forrester2009surrogate,brunton2020mlfluids}. A programmable CFD workflow could translate an engineering objective into a sequence of reusable operations and return simulation results with an interpretation of their significance. System response quantities, such as drag, lift, or pressure loss, could then guide subsequent evaluations within an active-learning or optimization loop, as illustrated by recent work on language-model-assisted CFD optimization \cite{chen2026optmeta}.

Large language models (LLMs) offer a way to coordinate such operations through natural-language interpretation, reasoning, and tool use \cite{yao2023react,schick2023toolformer}, supported by retrieval of technical information \cite{lewis2020rag}. OpenFOAM \cite{weller1998tensorial} provides an accessible, open-source CFD framework with a programmable interface that facilitates integration with LLM-based approaches. Several agents now generate and run OpenFOAM cases from natural language: MetaOpenFOAM and MetaOpenFOAM 2.0 \cite{chen2024metaopenfoam,chen2025metaopenfoam2}, OpenFOAMGPT and OpenFOAMGPT 2.0 \cite{pandey2025openfoamgpt,feng2026openfoamgpt2}, fine-tuned case generators \cite{dong2025autocfd}, Foam-Agent \cite{yue2026foamagent}, and ChatCFD \cite{fan2026chatcfd}. Their evaluations mainly measure whether a generated case executes and how closely its configuration or its computed fields match a reference. A run is treated as successful when the solver finishes without error or reports convergence, or when an LLM or a human reviewer judges the result. Language-guided multi-agent automation has also been applied to complex flow simulation \cite{xu2025cfdagent}. Two recent systems address physical correctness more directly. PhyNiKCE separates neural planning from symbolic constraint enforcement: before the LLM writes the case files, it retrieves configuration templates, boundary-condition knowledge, and solver constraints that are consistent with the requested physics \cite{fan2026phynikce}. AI CFD Scientist adds checks after execution, a mesh-independence gate and a vision-language physics check, and it evaluates the visual check on deliberately planted failures \cite{somasekharan2026aicfd}.

In this work, we introduce CFD Forge, an agentic framework for OpenFOAM built on one central idea: the language model should not decide whether a CFD result is acceptable. Most of the CFD agents described in the previous paragraph treat a run that executes without error as a success and score its accuracy afterwards, or leave the judgment to a language model or a person; AI CFD Scientist adds a mesh-independence gate, but its physics verdict still comes from a vision-language model (Table~\ref{tab:related}). CFD Forge instead makes acceptance a deterministic, physics-based decision. Each kind of flow it handles, such as the nozzle or the forward-facing step, is registered as a flow family with its own acceptance contract: a fixed set of numerical checks, including tests for flows that have not yet become statistically steady and for mass conservation in transient runs. Only that contract can return \texttt{ACCEPT}, \texttt{REJECT}, or \texttt{INCONCLUSIVE}. The model does the open-ended work, but its corrections come from a closed list and a validator can refuse them, including a premature \texttt{ACCEPT}. We test this split directly, using planted faults that a model judging alone accepted almost every time. Starting from a plain-language engineering request, CFD Forge builds the case, generates the mesh, runs the solver, diagnoses the result, and reports what it found, using an LLM to plan and propose while deterministic code carries out each step and checks the result. This division of labor builds on simulation agents developed for structural design \cite{jadhav2026mechdesigner}, protein molecular dynamics (MD) \cite{baratifarimani2025namdagent}, and adsorption calculations \cite{ock2026adsorbagent}. The closest precedent is PolyJarvis \cite{zhao2026polyjarvis}, which executes all-atom polymer MD using a validated run plan, deterministic stage scripts, and equilibration gates. It invokes a recovery agent only when structured failures occur, with a fixed decision budget and a closed set of recovery actions. CFD Forge extends this approach to CFD, including validator-controlled mesh refinement. Whereas MeshDQN uses reinforcement learning to coarsen a mesh while preserving a target quantity \cite{lorsung2023meshdqn}, CFD Forge restricts mesh changes to bounded refinements that follow a registered rule, are proposed by the model, and require validator approval.

CFD Forge thus occupies a complementary design point. Separating model proposals from deterministic checks is not new, and we do not claim it as such. What this paper contributes is how that separation is built and tested for CFD acceptance, summarized at the end of this section.

\begin{table}[!htbp]
\centering
\caption{Acceptance-related features described in prior agents and in CFD
Forge. Entries summarize each paper as we read it; \yes{} means the feature is described; ``--'' means the feature
is not described there, not that the system cannot have it. Several of these
systems cover far more cases and solver configurations than the three studies
reported here, and none has been compared with CFD Forge on common tasks.
PolyJarvis is a molecular-dynamics agent and is included as the closest
architectural precedent.}
\label{tab:related}
\scriptsize
\setlength{\tabcolsep}{2.5pt}
\begin{tabularx}{\linewidth}{@{}>{\raggedright\arraybackslash\hspace{0pt}}p{2.25cm}
  >{\raggedright\arraybackslash\hspace{0pt}}X >{\raggedright\arraybackslash\hspace{0pt}}p{1.5cm}
  >{\raggedright\arraybackslash\hspace{0pt}}p{1.65cm} >{\raggedright\arraybackslash\hspace{0pt}}p{1.45cm}
  >{\raggedright\arraybackslash\hspace{0pt}}p{1.25cm} >{\raggedright\arraybackslash\hspace{0pt}}p{2.0cm}@{}}
\toprule
System & Acceptance decided by & Registered numerical criteria & Bounded actions with deterministic veto & Development or stationarity test & Transient mass balance & Evaluation reported \\
\midrule
MetaOpenFOAM 2.0~\cite{chen2025metaopenfoam2} & Execution checks, then LLM reviewer, then human judgment & -- & -- & -- & -- & 13 requests, 10 runs each, ablation \\
OpenFOAMGPT 2.0~\cite{feng2026openfoamgpt2} & Error-free completion & -- & -- & -- & -- & Case studies, over 450 simulations \\
Foam-Agent~\cite{yue2026foamagent} & Error-free execution; field-level fidelity scored afterwards & -- & -- & -- & -- & 110 basic and 16 advanced tasks with field-level fidelity \\
ChatCFD~\cite{fan2026chatcfd} & Error-free, converged execution; fidelity scored afterwards & -- & -- & -- & -- & 315 cases with fidelity scoring \\
PhyNiKCE~\cite{fan2026phynikce} & Execution without failure; accuracy scored afterwards & -- & -- & -- & -- & 13 setups, repeated runs, ablation \\
AI CFD Scientist~\cite{somasekharan2026aicfd} & Vision-language verdict with a mesh-independence flag & Partial (mesh threshold only) & -- (rerun controller revises the request) & -- & -- & 5 tasks; 16 planted failures \\
PolyJarvis~\cite{zhao2026polyjarvis} (MD) & Deterministic gates & \yes & \yes{} (recovery set, fixed budget) & \yes{} (drift, block error) & n/a & 7 polymers, 3 replicates, vs.\ experiment \\
\midrule
CFD Forge & Deterministic family contract & \yes & \yes{} (per-family vocabulary) & \yes{} (nozzle monitors and fields; cube force components) & \yes & 2 agent families (19 sessions) plus one offline cube study; controller comparison, repeated calls \\
\bottomrule
\end{tabularx}
\end{table}

Automating these steps is useful only if the results can be trusted, but in CFD the usual indicators of a completed run do not establish that the result is correct. A solver can terminate normally, residuals can decrease, and the fields can appear physically reasonable while the boundary conditions are wrong, discretization error remains significant, or the flow has not yet reached a statistically steady state. For this reason, verification and validation rely on several independent kinds of evidence \cite{oberkampf2002vv,celik2008discretization}. A language model asked to judge such a run can be misled by the same indicators: in our controller comparison, a model given the evidence without the check results proposed acceptance for 12 of the 13 detectable planted faults.

CFD Forge is built around this observation. In practice, the language model turns a request into a case specification, reads solver logs, conservation measures, and flow histories, proposes what to try next, and writes the engineering report. The model is deliberately not used to decide acceptance, where a rule suffices (Sec.~\ref{sec:controller}), and new flow families are added through their own contracts. In this paper, geometry enters through parameterized families; an earlier CAD$\rightarrow$Gmsh prototype is described in Appendix~\ref{app:repro}.

We demonstrate the workflow with a compressible nozzle assessed against quasi-one-dimensional theory and a forward-facing step assessed for transient numerical health and storage-aware mass conservation (Appendix~\ref{sec:step_setup}), including new cases that the agent built from text requests with different geometry and operating conditions. A three-dimensional turbulent cube provides a diagnostic stress test. Although execution completes and mean drag becomes comparatively stable, quantitative force and wake diagnostics reveal continuing lateral-mode growth. Applied retrospectively, a check for statistically steady flow withholds acceptance on this evidence, illustrating why several physical signals must be examined together.

The contributions of this paper are:
\begin{itemize}
    \item \textbf{An agentic workflow from request to report:} A language model turns a plain-language request into an executable OpenFOAM case within a registered family, diagnoses each run, proposes the next step, and writes the report. We show this in 19 sessions on a nozzle and a forward-facing step, including new cases built from text requests.
    \item \textbf{Acceptance decided by contracts, not by the model:} Registered numerical checks for each flow family alone decide the verdict. The model can only choose corrections from a closed list, and a deterministic validator can refuse them. For example, in one session on the Mach-3 forward-facing step, the request asked whether the result depended on the mesh. The model refined the mesh and, because both solutions passed their checks, proposed \texttt{ACCEPT}. But passing the checks shows that each solution is healthy, not that the answer has stopped changing with the mesh, and no limit had been set for how much change is acceptable. The validator therefore refused and returned \texttt{INCONCLUSIVE} rather than an unsupported claim. This is one of three recorded refusals.
    \item \textbf{Testing for results that appear complete but are not:} An automated workflow has to decide when a simulation is done. If a run terminates, a mean settles, or a result agrees with theory, this does not necessarily mean that the simulation is finished or consistent. Hence, the contracts check the quantity that would reveal an unfinished or inconsistent result. Examples include a settled mean drag, which is not enough to certify a cube flow with growing lateral force; agreement with theory, which is not enough to certify a nozzle run that has not settled; and transient mass conservation, which holds only when the change in stored mass is counted.
    \item \textbf{Evidence that the split matters:} In a controller comparison, the model judging alone accepted nearly all detectable planted faults, whereas the CFD Forge configuration accepted none. The two configurations also differ in prompting, and a fixed rule reached the same verdicts, so the model's value lies in the open-ended work rather than in the verdict.
\end{itemize}
The model automates the work; the contracts decide what can be trusted. Together they move CFD a step closer to the automated, repeatable evaluations that high-throughput design and active-learning loops require.

\section{The CFD Forge Agent}
\label{sec:framework}

CFD Forge organizes CFD analysis as a programmable workflow, driven by a language-model agent, that connects an engineering request to simulation execution, assessment, and reporting. The agent works through programmatic case-construction primitives and explicit scientific checks. Its central design principle is to separate proposing an interpretation or action from determining whether the resulting simulation satisfies the prescribed acceptance criteria.

\subsection{Workflow Overview}
\label{sec:workflow}

The workflow begins with a natural-language engineering request describing the intended flow problem and relevant geometry or operating parameters. The request is interpreted and matched to a registered simulation family. Each family supplies a scientific contract defining the physical assumptions, admissible inputs, numerical recipe, required diagnostics, and permitted corrective actions.

\begin{figure}[!htbp]
    \centering
    \includegraphics[width=\linewidth]{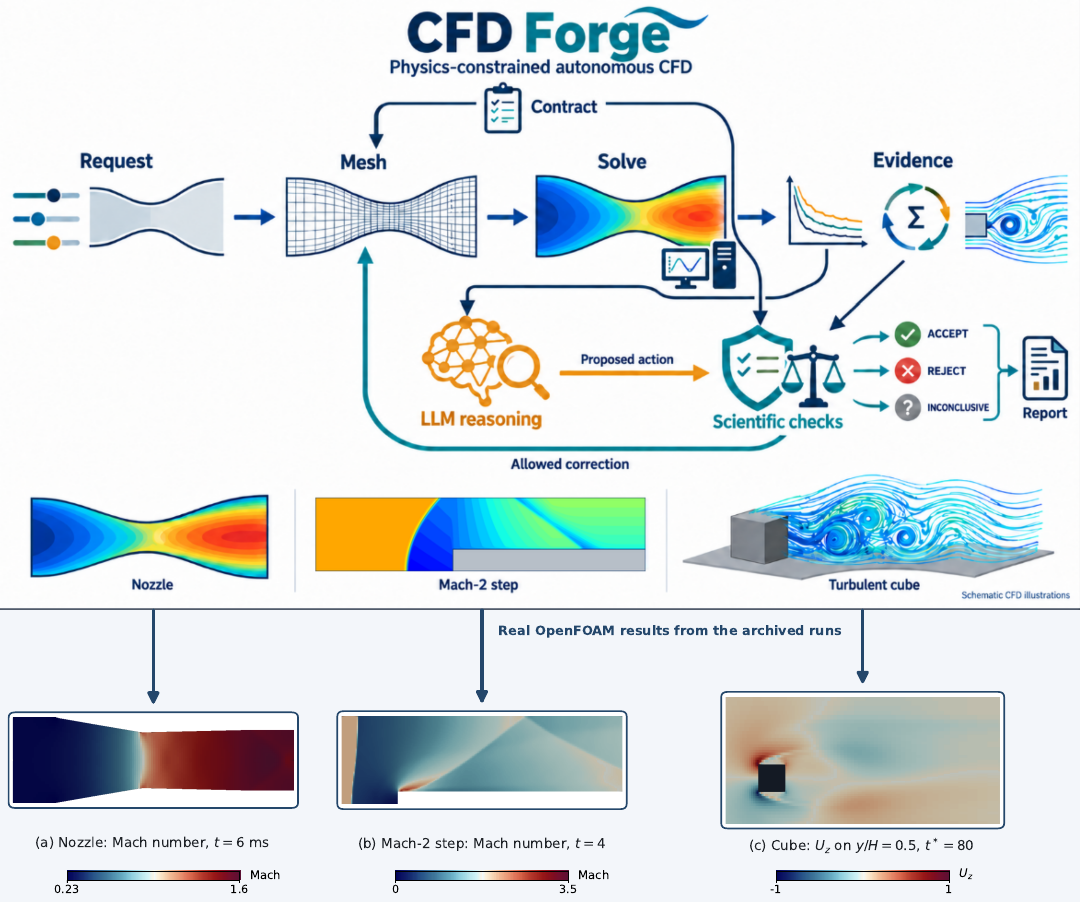}
    \caption{The CFD Forge workflow. Top: conceptual illustration (AI-generated, with
    the step tile redrawn by the authors; its flow pictures are schematic, not
    simulation output). A language model
    interprets the request and the simulation evidence and proposes actions;
    registered scientific checks, defined by each family's contract, decide
    \texttt{ACCEPT}, \texttt{REJECT}, or \texttt{INCONCLUSIVE} and permit only
    listed corrections. Bottom: the corresponding real OpenFOAM fields from the
    archived runs, each with its own color scale. (a) Nozzle and (b) Mach-2
    forward-facing step are from agent sessions; (c) the turbulent cube is a
    diagnostic study run outside the agent loop (Sec.~\ref{sec:cube_results}).}
    \label{fig:flowpilot_architecture}
\end{figure}

For an admissible request, a family-specific runner constructs the case, generates the mesh, assigns boundary and initial conditions, and executes OpenFOAM (Fig.~\ref{fig:flowpilot_architecture}). Diagnostic routines then extract evidence from the mesh, solver logs, saved fields, and time histories. This evidence supports both model-assisted interpretation and deterministic scientific assessment. Where an allowed corrective action is justified, the runner can update the permitted settings or continue execution before reassessing the result. Otherwise, the workflow terminates with a recorded decision and supporting artifacts.

\subsection{Reasoning and Scientific Authority}
\label{sec:authority}

Scientific acceptance is evaluated separately from the model's reasoning:
deterministic routines compute the registered quantities and compare them
with the criteria, and nothing the model says can override a failed or
unresolved check.

Within these bounds, the language model is the component that drives the
workflow by proposing what to do. It translates a natural-language engineering request
into a structured case specification that drives geometry, mesh, and
boundary-condition construction; reviews mesh evidence before setup proceeds;
diagnoses each completed run from solver, conservation, and field evidence;
and proposes the next action---accept, continue, extend the horizon, refine the
mesh, or stop safely---which the orchestrator executes, once approved, by invoking the
corresponding meshing, solver, and post-processing tools. It also interprets
rendered field images and writes the engineering summary.

The deterministic layer returns one of four decisions. The model never
writes the binding verdict: an \texttt{ACCEPT} from the model is an action
proposal, which the validators may refuse, whereas the decisions below are
computed by code:
\begin{itemize}
    \item \texttt{ACCEPT}: every registered check of the family passes, and
    every criterion the request depends on is registered.
    \item \texttt{CORRECT\_AND\_RERUN} (not terminal): the solution is healthy
    but incomplete---not yet stationary, short of the registered horizon, or
    insufficiently resolved---and an approved corrective action is executed
    before reassessment.
    \item \texttt{REJECT}: a registered criterion is violated by a solution
    that is not merely incomplete (a numerical failure, a boundary anomaly, or
    a request outside the family envelope), or a development test still fails
    when no permitted correction remains.
    \item \texttt{INCONCLUSIVE}: the assessment cannot be settled, because a
    criterion the request depends on is not registered, a needed correction
    was not approved, or the evidence is insufficient or anomalous.
\end{itemize}
Insufficient development is therefore not in itself a rejection. While a
permitted continuation exists, it leads to \texttt{CORRECT\_AND\_RERUN}; it
becomes \texttt{REJECT} only when the run ends still failing a registered
development test, as for the cube (Sec.~\ref{sec:cube_results}). A refused
action is also distinct from a rejected solution. The action validator rules
on the model's proposal, and the scientific validator rules on the result,
so a run can end with a healthy solution whose acceptance was refused
(Sec.~\ref{sec:agent_results}). The archived campaigns record
family-specific status strings, such as \texttt{PASS\_SINGLE\_MESH},
\texttt{FAIL}, and \texttt{STOPPED\_ACTION\_REFUSED}; the session ledger in
Appendix~\ref{app:repro} maps each to these decisions.

Acceptance therefore refers to the stated scientific contract and
the evidence available for that run. It does not imply that every
source of modeling or numerical uncertainty has been eliminated.

\subsection{Model Integration}
\label{sec:model}

\begin{figure}[!htbp]
    \centering
    \includegraphics[width=0.95\linewidth]{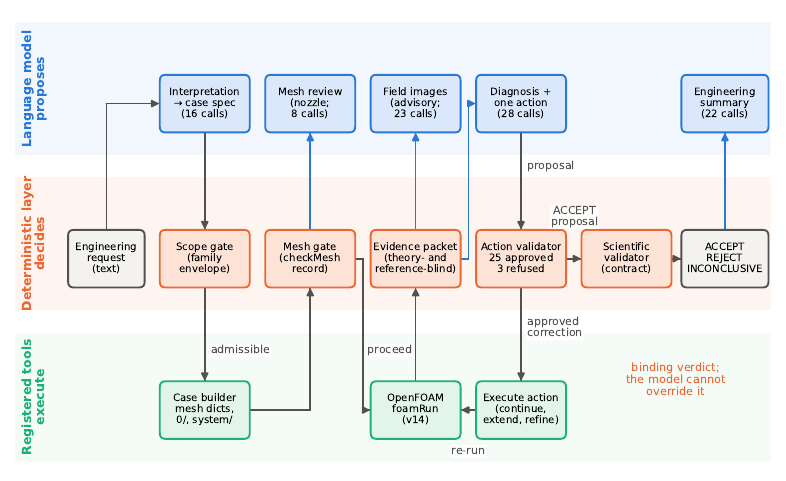}
    \caption{The agent loop as implemented. Blue: model calls, each a single
    request, schema-constrained except the summary; orange: deterministic gates and validators;
    green: registered tools. A proposed correction is executed only after the
    action validator approves it, and the run then returns to the solver; a
    proposed \texttt{ACCEPT} is passed to the scientific validator, whose
    verdict is binding. Call counts and rulings are recorded totals over the 19
    archived nozzle and step sessions (Sec.~\ref{sec:agent_results}). Mesh
    review is used in the nozzle family; field observation is advisory and
    feeds no gate.}
    \label{fig:agent_loop}
\end{figure}

Figure~\ref{fig:agent_loop} shows how the model calls sit inside the
execution loop, and Table~\ref{tab:llm_stages} lists them.
Each call is a single request to the Google Gemini API through the
\texttt{google-genai} Python SDK. The request has a fixed system
instruction, a JSON evidence payload, and, except for the plain-text summary, a Pydantic
response schema; it uses
temperature 0, and the structured response is parsed into a typed record
before any code acts on it. The archived campaigns used
\texttt{gemini-3.5-flash-lite} for interpretation, mesh review, diagnosis,
and reporting, and the identifier is recorded with every logged call. If the model is unavailable or its response fails the schema, the stage
stops, unless a deterministic fallback has been explicitly enabled; any such
fallback is recorded as non-LLM in the call log. The fallback is enabled by
default only for the step report summary, and no fallback occurred in the
archived sessions. Each call is logged with its
stage, model identifier, latency, number of attempts, and parsed output. In
the archived sessions one repair call was not logged
(Sec.~\ref{sec:agent_results}); the released code records it.
The runs also use a multimodal observer on rendered field images. It does not
write the model identifier into its own record. We infer its model from the provider's
daily per-model usage records, which are included in the data deposit: its 15 calls in the step sessions went to
\texttt{gemini-3.6-flash}, its code default when the model variable is
unset, and no \texttt{gemini-3.6-flash} requests were made on the day
of the nozzle sessions, so its 8 nozzle calls used
\texttt{gemini-3.5-flash-lite}. Five of its 23 calls, all in the step
sessions, failed with an HTTP 503 (service
unavailable) error from the provider and were not retried; the other 18
returned a schema-valid observation. Observation is advisory, so no failure
changed a decision. System instructions, response schemas, and iteration
budgets are listed in Appendix~\ref{app:repro}.

\begin{table}[!htbp]
\centering
\caption{Model calls in the CFD Forge workflow. ``Consumer'' is the
deterministic component that receives the structured output.}
\label{tab:llm_stages}
\footnotesize
\setlength{\tabcolsep}{3pt}
\begin{tabularx}{\linewidth}{@{}>{\raggedright\arraybackslash}p{2.3cm}
  >{\raggedright\arraybackslash}X >{\raggedright\arraybackslash}X
  >{\raggedright\arraybackslash}p{3.1cm}@{}}
\toprule
Stage & Model input & Structured output & Consumer \\
\midrule
Request interpretation & Engineering request text & Case parameters and unsupported-physics flags (step: unstated fields returned empty and filled by family defaults) & Scope gate (family envelope) \\
Mesh-evidence review (nozzle) & checkMesh and mesh-generation records & Proceed or stop, with cited evidence & Mesh gate \\
Diagnosis and action & Theory- or reference-blind evidence packet: solver health, conservation, stationarity, admissibility, boundaries, mesh, flow features & Diagnosis label, cited evidence, competing hypotheses with evidence for and against (nozzle), one action from the family vocabulary, confidence & Action validator, then scientific validator \\
Field observation & Rendered field images & Qualitative observations and suspected regions & Advisory only; no gate \\
Reporting & Final evidence and verdict & Engineering summary & Report \\
\bottomrule
\end{tabularx}
\end{table}

The diagnosis stage is the one that drives the workflow. Its system
instruction states that OpenFOAM solves the equations, that deterministic
code measures the evidence, and that the model interprets the evidence and
chooses one permitted action. It also directs the model not to invent
measurements, to distinguish insufficient run time from local
under-resolution and from numerical failure, not to treat solver completion
as evidence of stationarity, and to request a diagnostic when the evidence
cannot separate competing hypotheses. Analytical and reference targets are
removed from the evidence packet before the call, and a key-based check
rejects a packet that still contains them. The model therefore diagnoses from
the simulation's own evidence, and agreement with theory is checked
afterwards, by code.

Actions come from closed, family-specific vocabularies (listed in
Appendix~\ref{app:repro}). The action validator checks each proposal against
the evidence and a fixed iteration budget: for example, \texttt{CONTINUE\_RUN}
is refused once the requested horizon has been reached, and, for the step
family, \texttt{ACCEPT} is refused unless the registered scientific status
is final; for both families the final decision is taken by the scientific
validator.

\subsection{Scientific Contracts and Corrective Actions}
\label{sec:contracts}

A scientific contract makes the basis for assessment explicit before
a result is accepted. It specifies the geometry and operating range,
physical model, boundary-condition formulation, mesh requirements,
numerical controls, diagnostic quantities, and reference comparisons
applicable to a family. The criteria reflect the intended flow
regime: a steady nozzle requires appropriate stationarity evidence,
whereas an unsteady compressible step requires conservation checks
that account for changes in stored mass.

Corrective actions are restricted to those supported by the registered
workflow. The model proposes an action, while deterministic checks
establish whether it is admissible and whether another solver attempt
is warranted. The resulting case is then assessed using the same
acceptance criteria rather than a revised threshold chosen to match
the outcome.

\subsection{Evidence, Reporting, and Reproducibility}
\label{sec:provenance}

CFD Forge retains the artifacts needed to inspect both the simulation
and the decisions made around it: case configurations, mesh information,
solver logs, diagnostic measurements, model-call records where available,
proposed actions, and final assessments, together with plots, contours, and
field-evolution videos rendered from the saved fields. Together they give an
auditable account of what was requested, what was computed, which actions
were taken, and why the result received its final decision.

\section{Walkthrough: From an Engineering Request to an Accepted Simulation}
\label{sec:walkthrough}

We first follow one complete agent session, the canonical Mach-2
forward-facing-step run, from the engineering request to the final verdict.
Every artifact quoted below is taken from the archived session record.
Table~\ref{tab:timeline} in Appendix~\ref{app:records} lists the stages, who performed each one, and when.

\paragraph{Request.} The session starts from a short natural-language request:

\begin{quote}\small\itshape
Simulate a 2D forward-facing step in the validated supersonic channel family
with inlet Mach number 2.0, step height 0.20 channel heights, and step location
x = 0.6. Run the CFD, inspect the transient shock/compression structure, and
determine whether the final result passes deterministic scientific validation.
\end{quote}

\paragraph{Interpretation.} The model converts the request into a structured
specification (Fig.~\ref{fig:walk_spec} in Appendix~\ref{app:records}). It extracts the three quantities
the user stated---Mach number 2.0, step height 0.2, and step location
0.6---and leaves every quantity the user did not state empty rather than
guessing it. It also records the two reporting objectives separately from
the case parameters. The framework then fills each empty field from the
family defaults (the Foundation v14 \texttt{shockFluid/forwardStep} recipe)
and labels every field as \emph{stated by the user} or \emph{inferred from
family defaults}, so the origin of each simulation parameter is explicit.

\paragraph{Admissibility and case construction.} A deterministic scope gate
checks the resolved specification against the family envelope: Mach number,
step height and position as fractions of the channel, channel aspect ratio,
cell count and aspect ratio, thermodynamic state, end time, and Courant
limit. All fourteen checks pass, and the gate records the realized inlet
Mach number, 2.000006. The family builder then writes a three-block
\texttt{blockMeshDict} ($240\times80$ cells minus the step, 16,128 cells,
$\Delta x=\Delta y=0.0125$) and sets the inlet and initial state
$p=T=1$, $\mathbf{U}=(2,0,0)$. It copies the numerical recipe unchanged and
hash-checks it, and runs \texttt{blockMesh} and \texttt{checkMesh}, which
reports a valid mesh with two solution directions.

\paragraph{Execution and evidence.} OpenFOAM \texttt{shockFluid} advances
the flow to $t=4$ in 5,428 steps (102\,s). Deterministic diagnostics then
compute the 17 hard checks and the compression measurements, and render
eight field images. The evidence packet passed to the model contains solver
health, horizon, mesh quality, field ranges, Courant history, the transient
mass balance, the measured compression structure, the outcome of each
deterministic check, the iteration budget, and the permitted diagnoses and
actions. Reference comparisons are removed before the call. In this family
the model therefore sees the check outcomes; its task is to interpret them
and to choose the next action, not to recompute them.

\paragraph{Diagnosis and decision.} A multimodal LLM observer first
inspects the field images. The diagnosis call then returns
\texttt{HEALTHY\_COMPLETE} with high confidence, cites solver health,
horizon, mesh, field ranges, Courant number, transient conservation, and the
deterministic checks, and proposes \texttt{ACCEPT}. The action validator
approves the proposal because it is consistent with the registered checks,
and the scientific validator independently returns
\texttt{PASS\_2D\_FORWARD\_STEP}. The terminal verdict is \texttt{ACCEPT}.

\paragraph{Report.} Finally the model writes the engineering summary from
the evidence and the verdict. It reproduces the measured values correctly
but attaches physical units (s, Pa, K) to the normalized quantities and adds
a sentence about quasi-one-dimensional nozzle comparisons that does not apply
to this case. Such errors are why the summary is treated as narrative and
never as evidence: every number in it can be traced back to, and checked
against, the deterministic record.

The full session took 165\,s from request to report. The OpenFOAM run took
102\,s of that, and the four model calls took about 22\,s.

\section{Simulation Families and Assessment Protocol}
\label{sec:experiments}

We evaluate the agent on three CFD studies whose acceptance logic differs.
The converging--diverging nozzle is a steady compressible flow with an
analytical reference. The Mach-2 forward-facing step is a transient,
shock-containing flow whose acceptance rests on storage-aware conservation
and compression diagnostics. The three-dimensional turbulent cube tests
whether the acceptance layer can distinguish numerical completion from a
developed flow. All three use OpenFOAM Foundation v14.
Figure~\ref{fig:case_geometries}, Fig.~\ref{fig:meshes_step_cube} and
Table~\ref{tab:setup} (the latter two in Appendix~\ref{app:setup}) summarize their
geometry, meshes, boundary conditions, and numerics; the complete setup and
acceptance criteria are given in Appendix~\ref{app:setup}.

The nozzle and step runs were executed by the agent from natural-language
requests, and their model calls are recorded, except for one repair call
disclosed in Sec.~\ref{sec:agent_results}. The cube campaign was executed
outside the agent loop; afterwards we applied a diagnosis call in the style of
the agent's diagnosis stage to its archived evidence three times and ruled on each proposal with the
stationarity gate, which was registered retrospectively, after the cube
data existed (Sec.~\ref{sec:cube_results}, Appendix~\ref{app:repro}). Archived-case replay is used only to check
that recorded verdicts are reproduced, never as evidence of new execution.

\begin{figure}[!htbp]
    \centering
    \includegraphics[width=0.92\linewidth]{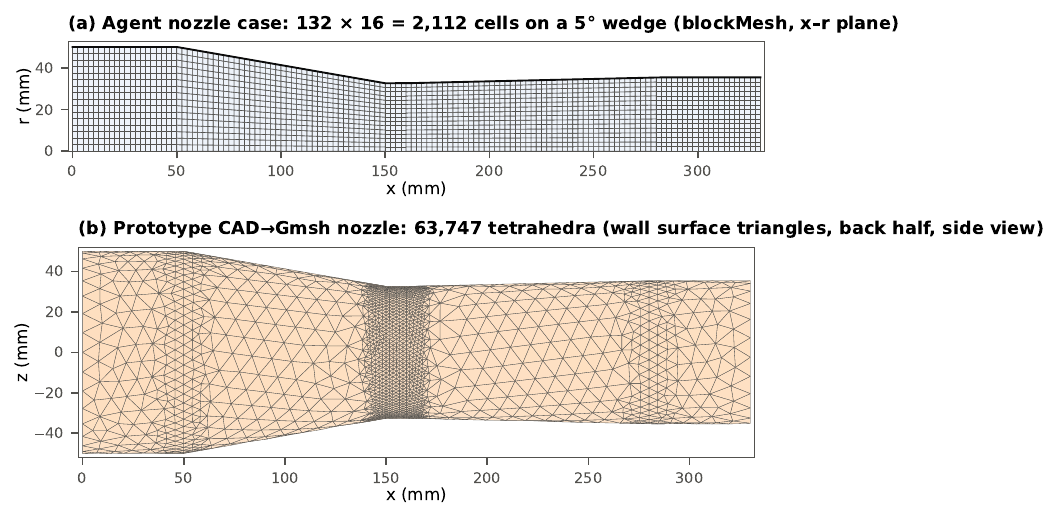}
    \caption{Nozzle meshes. (a) The agent-built wedge mesh in the $x$--$r$
    plane, every cell shown: five blocks with 20, 40, 4, 48, and 20 axial
    divisions and 16 radial divisions, reconstructed exactly from the archived
    \texttt{blockMeshDict} (linear edges, uniform grading), whose digest
    matches the native case. (b) The tetrahedral mesh from the earlier
    CAD$\rightarrow$Gmsh prototype of the same nozzle
    (Appendix~\ref{app:repro}): wall-surface triangles of the archived
    \texttt{.msh} file, back half only, viewed from the side, showing the
    throat refinement selected by the model.}
    \label{fig:case_geometries}
\end{figure}

\section{Results}
\label{sec:results}

\subsection{Agent Behavior Across Campaigns}
\label{sec:agent_results}

\begin{figure}[!htbp]
    \centering
    \includegraphics[width=\linewidth]{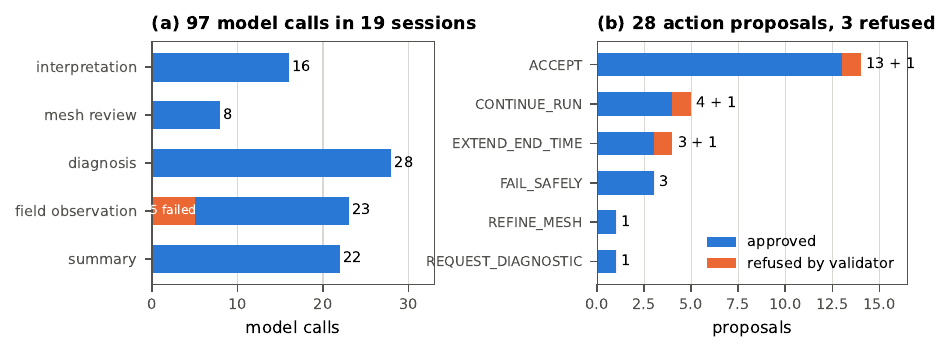}
    \caption{Model activity in the 19 archived agent sessions, counted from
    every call record in the session archive and deduplicated per run
    (Appendix~\ref{app:repro}); the one unrecorded repair call is not
    included. (a) Model calls by stage; 5 of the 23
    field-observation calls failed with a provider-side HTTP 503 error.
    (b) Action proposals by type and the action validator's ruling.}
    \label{fig:agent_stats}
\end{figure}

Table~\ref{tab:assessment} summarizes the registered outcomes. The archived
campaigns comprise 19 agent sessions (8 nozzle, 11 step). These are 16 runs,
because one step run was resumed across four sessions, and 11 distinct
requests, because the nozzle requests were repeated while the nozzle
pipeline was being developed (Appendix~\ref{app:repro}). Across them the call
records contain 97 model calls: 16 request interpretations, 8 mesh reviews, 28
diagnoses, 23 field observations, and 22 summaries. All 74 text calls
succeeded on the first attempt, with median latencies between 1.9 and 4.9\,s
per stage; 5 of the 23 field-observation calls failed with a provider error
(Sec.~\ref{sec:model}). Token counts were not logged in these campaigns. The
28 recorded diagnoses produced 28 recorded action proposals: 14 \texttt{ACCEPT}, 5
\texttt{CONTINUE\_RUN}, 4 \texttt{EXTEND\_END\_TIME}, 3
\texttt{FAIL\_SAFELY}, 1 \texttt{REFINE\_MESH}, and 1
\texttt{REQUEST\_DIAGNOSTIC}. The validator approved 25 and refused 3
(Fig.~\ref{fig:agent_stats}). Table~\ref{tab:actions}
lists representative decision points, and Fig.~\ref{fig:trace} in
Appendix~\ref{app:records} shows one complete diagnosis record.
These counts are taken from the call records. In one nozzle iteration (N4,
second iteration) the diagnosis module, after the action validator refused
its first proposal, made a second call with the refusal reasons and recorded
only the second proposal; the refused proposal and the extra call are
therefore not included in the counts above.

In the nozzle continuation, the model read the stationarity failures and
proposed \texttt{CONTINUE\_RUN}. In a Mach-3 step run that had reached only
$t=2$ of the registered $t=4$ horizon, it proposed extending the end time, a
bounded change ($2\to4$) that the validator permitted before reassessment and
acceptance. In a Mach-3 run with a taller step ($h=0.3$), no compression front
could be measured; the model attributed this to upstream migration of the
shock system (Sec.~\ref{sec:variants}) and proposed \texttt{FAIL\_SAFELY}. That run was preserved and
rejected rather than tuned.

The three recorded refusals show the validator correcting the model. In the
iterative Mach-3 run, the model proposed \texttt{CONTINUE\_RUN} after the
requested horizon had been reached, which was refused because more physical
time requires \texttt{EXTEND\_END\_TIME}; a later \texttt{EXTEND\_END\_TIME}
was refused because the iteration budget was exhausted. That campaign was
completed in supervised resumed sessions, after the restart-seam treatment in
the mass-balance diagnostic had been corrected, so its final acceptance is not
an unattended result. The clearest case is a Mach-3 request for a
numerical-sensitivity assessment. The model proposed \texttt{REFINE\_MESH},
which was approved under the registered bounded policy (linear refinement by a
factor of 2, from 4,032 to 16,128 cells, as a fresh solve). Both grids passed
their hard checks, and the model proposed \texttt{ACCEPT}. The validator
refused it, because no cross-grid tolerance is registered for the step family,
and the run ended as \texttt{STOPPED\_ACTION\_REFUSED}, which is
\texttt{INCONCLUSIVE} under Sec.~\ref{sec:authority}: both solutions were
healthy, but the requested sensitivity could not be certified.

One stop was caused by the deterministic layer itself. In the first step
session, the validator reported a fatal solver error because it matched the
OpenFOAM start-up line that enables floating-point trapping; the model, which
is instructed to treat the checks as authoritative, proposed
\texttt{FAIL\_SAFELY}, which was approved. After the detector was corrected, a
deterministic reanalysis of the unchanged run returned
\texttt{PASS\_2D\_FORWARD\_STEP} without a new solve. Deterministic authority
is therefore only as reliable as the checks it applies.

In every recorded case the terminal verdict was computed by the validators.
For the continuation, extension, and safe-stop decisions above, a rule keyed to
the failed check would have chosen the same action, and the five failed
observer calls changed no outcome.

\begin{table}[!htbp]
\centering
\caption{Registered scientific assessment of the three studies.}
\label{tab:assessment}
\footnotesize
\setlength{\tabcolsep}{4pt}
\begin{tabularx}{\linewidth}{@{}>{\raggedright\arraybackslash}p{2.3cm}
  >{\raggedright\arraybackslash}X >{\raggedright\arraybackslash}X
  >{\raggedright\arraybackslash}X@{}}
\toprule
Evidence & Nozzle & Forward-facing step & Cube \\
\midrule
Mesh & checkMesh OK & checkMesh OK; 2 solution directions & checkMesh OK \\
Solver health & Completed; $p,T,\rho>0$ at all saved times & Completed; positive at every step; no fatal signature & Completed; no warnings in final segments \\
Conservation & Window mismatch $0.055\%$ ($<0.1\%$) & Closure residual $5.7\times10^{-11}$ ($<10^{-6}$) & Volume-flux imbalance $3\times10^{-11}$ \\
Physical or reference & Canonical run: quasi-1-D differences $\le1.83\%$ & $p$, $\rho$, $T$ rise across front & Not assessed (no accepted reference data) \\
Stationarity or development & Drift $0.081\%$, field change $0.111\%$ ($<0.2\%$) & Not required (transient family) & Lateral growth ratio 2.10 ($>1.25$) \\
Registered outcome & \texttt{ACCEPT} \newline (\texttt{PASS\_SINGLE\_MESH}) & \texttt{ACCEPT} \newline (\texttt{PASS\_2D\_FORWARD\_STEP}) & \texttt{REJECT} \newline (\texttt{STILL\_DEVELOPING}; retrospective gate) \\
\bottomrule
\end{tabularx}
\end{table}

\begin{table}[!htbp]
\centering
\caption{Recorded decision points with genuine LLM calls.
``Approved'' and ``Refused'' denote rulings of the
deterministic action validator; the last column gives the archived family status computed
by the scientific validator, with the decision of Sec.~\ref{sec:authority}
in parentheses.}
\label{tab:actions}
\footnotesize
\setlength{\tabcolsep}{3pt}
\begin{tabularx}{\linewidth}{@{}>{\raggedright\arraybackslash}p{2.2cm}
  >{\raggedright\arraybackslash}X >{\raggedright\arraybackslash}p{2.55cm}
  >{\raggedright\arraybackslash}p{1.45cm} >{\raggedright\arraybackslash}p{3.3cm}@{}}
\toprule
Run and state & Deterministic evidence & Model proposal & Validator & Verdict \\
\midrule
Nozzle, $t=1$\,ms & 17/20 checks; mass mismatch $1.35\%$, drift $2.0\%$, field change $1.2\%$ & \texttt{CONTINUE\_RUN} & Approved & \texttt{FAIL} (\texttt{CORRECT\_AND\_\allowbreak RERUN}) \\
Nozzle, $t=6$\,ms & 20/20 checks & \texttt{ACCEPT} & Approved & \texttt{PASS\_SINGLE\_MESH} \\
Step, Mach 2, $t=4$ & 17/17 hard checks; compression rises measured & \texttt{ACCEPT} & Approved & \texttt{PASS\_2D\_FORWARD\_STEP} \\
Step, Mach 3, $t=2$ of 4 (resumed) & Hard checks pass; registered horizon not reached & \texttt{EXTEND\_END\_TIME} ($2\to4$) & Approved & Target not reached (\texttt{CORRECT\_AND\_\allowbreak RERUN}) \\
Step, Mach 3, $t=4$ & 17/17 hard checks & \texttt{ACCEPT} & Approved & \texttt{PASS\_2D\_FORWARD\_STEP} \\
Step, Mach 3, $h=0.3$ & Compression front not measurable & \texttt{FAIL\_SAFELY} & Approved & \texttt{FAIL} (preserved) \\
Step, Mach 3, sensitivity request, 4,032 cells & Hard checks pass; one grid cannot establish sensitivity & \texttt{REFINE\_MESH} ($\times2$) & Approved & \texttt{PASS} on one grid (\texttt{CORRECT\_AND\_\allowbreak RERUN}) \\
Same, 16,128 cells & Hard checks pass; no registered cross-grid tolerance & \texttt{ACCEPT} & \textbf{Refused} & \texttt{STOPPED\_ACTION\_\allowbreak REFUSED} (\texttt{INCONCLUSIVE}) \\
\bottomrule
\end{tabularx}
\end{table}

\subsection{Controller Comparison}
\label{sec:controller}

The session records above show that the interface behaves as designed; they
do not show what the deterministic layer contributes. We therefore replayed
archived decision points through four controllers using the same model
(\texttt{gemini-3.5-flash-lite}, temperature 0) and the current validators.
(A)~A fixed rule maps the failed checks to an action, with no model call.
(B)~CFD Forge: the model diagnoses the full evidence packet, including the
deterministic check results, and proposes an action; the action gate screens
the proposal and the validator decides. (B$'$)~Gates off: the same model
proposal is taken directly as the verdict, without the action gate or the
validator. (C)~Model only: the model returns a verdict
(\texttt{ACCEPT}, \texttt{NEEDS\_MORE\_RUNTIME}, \texttt{REJECT} or
\texttt{INCONCLUSIVE}) from a filtered diagnostic packet: the aggregate
validator outcomes, pass/fail summaries and failed-check lists are removed,
while the measured quantities, selected numerical-health and mesh flags (for
example \texttt{solver\_completed} or \texttt{mesh\_ok}), and a one-sentence
family description remain. Arm~C therefore tests whether the model, given
this filtered evidence, reaches the verdict the contract would reach. Arms
B/B$'$ and C also differ in system instruction, response schema and
decision vocabulary, so the comparison contrasts two configurations rather
than isolating the check results alone. For the cube, the diagnosis packet
given to arm~B also excludes the stationarity verdict and its threshold; there
the arms differ in prompting, output interpretation and enforcement, not in
access to the gate result. Each archived and defect point was queried five
times and each fault three times; arms B and B$'$ share one model call per
repeat.

The decision points are of three kinds. Sixteen archived points cover nine
forward-facing-step states, six nozzle states and the cube
(Table~\ref{tab:controller}). The reference answer for each is the
registered decision: \texttt{ACCEPT} when every check passes,
\texttt{CORRECT\_AND\_RERUN} when only the target time or settling checks
are unmet, and not-acceptable otherwise. Eighteen planted faults each alter
one diagnostic of an accepted Mach-2 step run, for example marking the solver
as not completed, setting a field minimum negative or adding a wall flux; none
should be accepted. Two defect points reproduce historical diagnostic-extraction
errors, a false-positive fatal-error flag and a pre-fix restart-seam closure,
under which healthy runs appear to fail.

\begin{table}[!htbp]
\centering
\caption{Controller comparison with \texttt{gemini-3.5-flash-lite}. Entries
are correct decisions / decisions, with false accepts (a not-acceptable state
accepted) in parentheses. A decision counts as correct when it equals the
registered decision; where the reference is only ``not acceptable'' (planted
faults and some archived states), any non-accepting decision counts as
correct. Differences in correct counts between B and B$'$ therefore reflect
the disposition (\texttt{REJECT} versus \texttt{INCONCLUSIVE}), not false
acceptance. Arms A and B terminate at the same validator and so
reach the same verdicts by construction; they differ in the actions proposed.
Of 288 model calls, four failed and are excluded.}
\label{tab:controller}
\small
\begin{tabular}{@{}lccc@{}}
\toprule
Controller & Archived (16 points) & Planted faults (18) & Defects (2) \\
\midrule
A: fixed rule & 16/16 (0) & 18/18 (0) & 0/2 (0) \\
B: CFD Forge & 79/79 (0) & 53/53 (0) & 0/10 (0) \\
B$'$: gates off & 74/79 (0) & 53/53 (0) & 0/10 (0) \\
C: model only, blind packet & 72/79 (7) & 3/53 (50) & 5/10 (0) \\
\bottomrule
\end{tabular}
\end{table}

Without the check results, the model-only arm (C) often proposed accepting
states that the contract rejects. It proposed accepting the Mach-3, $h=0.3$
step case in all five repeats, although its compression front could never be
measured, and the flawed nozzle continuation in two of five, giving different
answers to the same evidence. It did not, however, ask for any unnecessary
reruns.

The planted faults tell a clearer story. Each was queried three times, and
arm~C proposed acceptance for 17 of the 18 in every completed repeat (one call
failed, on a fault the model could not see); only the explicit fatal-error
flag stopped it. We call a fault detectable when the field it changes appears
in the model-only packet. Thirteen were detectable, such as
\texttt{solver\_completed: false}, and the model still proposed accepting 12
of them in all three repeats, usually reasoning that the run had reached its
target time of $t=4$. The other five (an edited recipe, a subsonic inlet,
initial-state drift, a missing boundary patch, and a cyclic patch) changed only
fields outside the packet, so their acceptance says more about the packet than
about the model.

With the deterministic results (arm~B$'$), the same model proposed acceptance
only when the checks passed: no false accept occurred in the 93 decisions on
states that should not be accepted. Its five errors on archived points are all
on the no-front step case, where it proposed \texttt{FAIL\_SAFELY}, which maps
to \texttt{INCONCLUSIVE} rather than \texttt{REJECT}. Because the model never
proposed acceptance contrary to a failed check, enforcement (arm~B) changed no
acceptance outcome relative to arm~B$'$. It changed the disposition in 66 of
the 142 decisions, turning 50 \texttt{INCONCLUSIVE} and 16
\texttt{CORRECT\_AND\_RERUN} outcomes into \texttt{REJECT}, and the action gate
refused two proposals, both \texttt{REFINE\_MESH} on faults whose failed checks
(\texttt{mesh\_ok}, \texttt{two\_solution\_directions}) are not resolution
problems. Enforcement of the registered checks held in every case but was not
exercised against a model that proposed an unjustified acceptance.

The defect points show the limit of every evidence-dependent controller. With
the false fatal-error flag, all arms withheld acceptance of a healthy run. With
the pre-fix seam closure, arms A, B and B$'$ stopped a run that needed only
more time, whereas arm~C, which did not see the faulty closure verdict,
correctly asked for more runtime. A contract protects against model error, not
against an error in its own diagnostics; here its failures were conservative,
and neither defect produced a false accept.

{\sloppy
Repeat consistency was also measured. Arm~B gave the same diagnosis and action
in all repeats at 16 of 16 archived points, including five of five on the cube
evidence, and at 13 of 18 faults; at the other five faults the proposed
corrective action varied (\texttt{FAIL\_SAFELY},
\texttt{REBUILD\_FROM\_VALIDATED\_SPEC}, \texttt{REFINE\_MESH},
\texttt{REJECT\_UNSUPPORTED}) but the verdict did not. The run made 288 model
calls, of which 284 succeeded, using $1.10\times10^6$ tokens at a median
latency of 1.27\,s per call. The script (\texttt{controller\_comparison.py} in
\texttt{paper/cfd\_forge/scripts/}), which also contains the prompts and the
list of blinded keys, and the per-call records under
\texttt{evidence/controller\_comparison/} are in the repository. The
comparison uses one small, low-latency model, archived rather than new runs,
and faults planted in one family; it is not a held-out benchmark, and a
stronger model may perform better in arm~C.

The comparison also bounds what the model contributes. At these decision
points the tested fixed-rule controller reached the same verdicts as CFD Forge,
and at every archived point it proposed the same action as the model. The two
differed only on planted faults, where the rule always failed safely and the
model, at 8 of 18 faults, proposed a correction other than a safe stop in at least one
repeat, such as \texttt{REDUCE\_MAX\_CO} for a non-finite Courant number or
\texttt{REBUILD\_FROM\_VALIDATED\_SPEC} for a cell-count mismatch; whether those
corrections would have succeeded was not tested. The model's role is therefore
not to improve acceptance decisions but to do the open-ended work that the
tested rule cannot: interpret a free-text request into a registered
specification, construct new cases within a family
(Sec.~\ref{sec:variants}), explain the evidence, and propose among several
admissible corrections. The comparison does not measure those contributions.
\par}
\subsection{New Cases Within the Registered Families}
\label{sec:variants}

Within each registered family, the agent constructed, ran, and assessed new
cases from natural-language requests that changed the geometry and the
operating conditions (Table~\ref{tab:variants}). In each case the model
extracted the stated quantities, the family builder produced a new mesh and
case, and the same contract assessed the result. Boundary-condition types are
fixed by the contract, so the requests changed boundary-condition values
($p_0$, the inlet Mach number), not their formulation; a request outside the
registered envelope would have been refused by the scope gate. These were
development runs, not a held-out evaluation (Appendix~\ref{app:repro}).

The results behave as the physics requires. Raising $p_0$ by $10\%$ at fixed
geometry left the exit Mach number unchanged (1.513) and scaled the exit
pressure and the mass flow by 1.100, as expected for a choked perfect-gas
nozzle; enlarging the exit raised the exit Mach number to 1.652, within
$0.39\%$ of quasi-one-dimensional theory. All three nozzle variants passed
the twenty registered checks, with exit Mach number, exit pressure, and mass
flow within $2\%$ of theory. In the
step family, the pressure rise across the detected front exceeded the
stationary normal-shock ratio at every Mach number by 4--15\%, which is
consistent with fronts that are still moving upstream, and the storage-aware
mass closure stayed between $3\times10^{-11}$ and $6\times10^{-11}$ in every
run. The quasi-one-dimensional choking estimate of Sec.~\ref{sec:step_results}
separates the configurations. Where the post-shock sonic area exceeds the
opening above the step (Mach 2 with $h=0.2$, and Mach 3 with $h=0.3$), the
front was displaced toward the inlet; with the step at $x=0.6$ and $h=0.3$
it reached $x=0.025$, no front could be measured, and the model proposed
\texttt{FAIL\_SAFELY}. With the same step moved to $x=1.0$, the front had not
yet reached the inlet at $t=4$ and the run passed its contract, which for
this transient family includes no stationarity criterion. The Mach-2.5 run
was first stopped by the false-positive fatal-error check of
Sec.~\ref{sec:agent_results} and passes after the corrected reanalysis.

\begin{table}[!htbp]
\centering
\caption{New cases built and run by the agent from text requests. Nozzle
values are at $t=6$\,ms on the 2,112-cell wedge, with differences from
quasi-one-dimensional theory in parentheses. Step values are at $t=4$ on the
family mesh ($\Delta x=\Delta y=0.0125$; the cell count varies with $h$
and $x_s$): the detected front position below the step height, the
pressure rise across it with the stationary normal-shock ratio in
parentheses, and the post-shock sonic area $A^*/A$ compared with the opening
$1-h$ above the step (choked when $A^*/A>1-h$).}
\label{tab:variants}
\footnotesize
\setlength{\tabcolsep}{3pt}
\begin{tabular}{@{}llll@{}}
\toprule
\multicolumn{4}{@{}l}{\emph{Converging--diverging nozzle}} \\
Request & Exit $M$ & Exit $p$ (kPa) & $\dot m$ (kg\,s$^{-1}$) \\
\midrule
Reference: $r_e=35.4$\,mm, $p_0=200$\,kPa & 1.513 ($+0.59\%$) & 53.14 ($-1.83\%$) & 1.541 ($-1.11\%$) \\
Exit radius $r_e=37.0$\,mm & 1.652 ($+0.39\%$) & 43.29 ($-1.57\%$) & 1.541 ($-1.10\%$) \\
Inlet total pressure $p_0=220$\,kPa & 1.513 ($+0.58\%$) & 58.46 ($-1.83\%$) & 1.695 ($-1.11\%$) \\
\midrule
\multicolumn{4}{@{}l}{\emph{Forward-facing step}} \\
Request ($M$, $h$, $x_s$) $\rightarrow$ decision & Front $x$ & $p$ rise (normal shock) & $A^*/A$ vs.\ $1-h$ \\
\midrule
2.0, 0.20, 0.6 $\rightarrow$ \texttt{ACCEPT} & 0.113 & 5.18 (4.50) & 0.822 $>$ 0.80, choked \\
2.5, 0.15, 0.6 $\rightarrow$ archived \texttt{REJECT}$^{a}$ & 0.341 & 7.49 (7.12) & 0.760 $<$ 0.85 \\
\quad corrected reanalysis $\rightarrow$ \texttt{PASS}$^{a}$ & & & \\
3.0, 0.10, 0.6 $\rightarrow$ \texttt{ACCEPT} & 0.450 & 11.11 (10.33) & 0.719 $<$ 0.90 \\
3.0, 0.20, 0.6 $\rightarrow$ \texttt{ACCEPT}$^{b}$ & 0.313 & 10.82 (10.33) & 0.719 $<$ 0.80 \\
3.0, 0.30, 0.6 $\rightarrow$ \texttt{REJECT} & 0.025 & not measurable & 0.719 $>$ 0.70, choked \\
3.0, 0.30, 1.0 $\rightarrow$ \texttt{ACCEPT} & 0.425 & 11.00 (10.33) & 0.719 $>$ 0.70, choked \\
3.5, 0.20, 0.6 $\rightarrow$ \texttt{ACCEPT} & 0.338 & 14.76 (14.12) & 0.692 $<$ 0.80 \\
\bottomrule
\multicolumn{4}{@{}p{0.97\linewidth}@{}}{\footnotesize $^{a}$False-positive fatal-error check; \texttt{PASS\_2D\_FORWARD\_STEP} on deterministic reanalysis after the fix. $^{b}$Reached after resumed sessions (Table~\ref{tab:ledger}); the mesh-sensitivity request with the same parameters ended \texttt{INCONCLUSIVE}.} \\
\end{tabular}
\end{table}

\subsection{Converging--Diverging Nozzle}
\label{sec:nozzle_results}

\begin{figure}[!htbp]
    \centering
    \includegraphics[width=\linewidth]{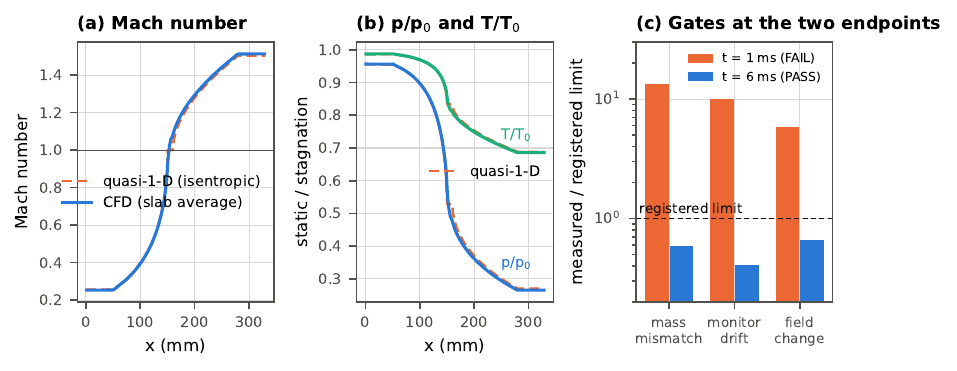}
    \caption{Nozzle results. (a,b) Axial profiles of the canonical run at
    $t=6$\,ms (solid) and the isentropic quasi-one-dimensional solution for
    the registered area distribution (dashed). (c) The three checks that failed
    at 1\,ms in the separate continuation run (mass balance, monitor drift, field change), expressed as measured value over
    registered limit, at the initial 1\,ms endpoint (failed) and after the
    approved continuation to 6\,ms (passed).}
    \label{fig:nozzle}
\end{figure}

The canonical nozzle run completed to $t=6$\,ms in 30,471 adaptive steps
(maximum Courant number 0.400) and passed all twenty registered checks,
yielding \texttt{PASS\_SINGLE\_MESH} and a deterministic \texttt{ACCEPT}.
Figure~\ref{fig:nozzle} compares radially volume-weighted axial profiles
with the isentropic quasi-one-dimensional solution for the same area
distribution. The flow accelerates from $M=0.253$ in the inlet section,
reaches $M=1.029$ in the throat, and leaves the nozzle at $M=1.513$, while
static pressure and temperature fall to $0.266\,p_0$ and $0.686\,T_0$. The
minimum slab-averaged Mach number downstream of the throat is 1.17 and the
minimum face-normal outlet Mach number is 1.46, so no internal normal shock
forms and the entire exit plane is supersonic. The exit static pressure,
53.1\,kPa, exceeds the 30\,kPa nominal ambient pressure, which is consistent
with an underexpanded nozzle.

The agent-constructed canonical run reproduces the coarsest mesh of the
LLM-free reference campaign exactly: every reported quantity, the 30,471
time steps, and the stationarity metrics agree to machine precision. The
reference refinement study (Table~\ref{tab:nozzle_grid}, Appendix~\ref{app:setup}) passed its
registered criteria, with last-pair changes of at most $0.41\%$ (throat
Mach number) and $0.27\%$ (exit pressure). Refinement increases the exit Mach
number from 1.513 to 1.524 and moves the solution slightly further from
quasi-one-dimensional theory, to $-3.16\%$ in exit pressure on the finest
mesh, which is still within the $5\%$ limit. This trend may reflect multidimensional corner effects that the
one-dimensional model omits; we have not established the cause. No formal grid-convergence index is reported.

Table~\ref{tab:nozzle} lists the deterministic comparison with theory.
The percentage differences for the exit Mach number, exit static pressure,
exit temperature, exit velocity, and mass flow are all below $2\%$; the
largest is the exit static pressure ($-1.83\%$, limit $5\%$), and the mass
flow is $1.11\%$ below the ideal choked value. The throat Mach number, 1.029,
is assessed against its registered band ($0.9<M_t<1.1$) rather than as a
percentage difference. These differences are consistent with a two-dimensional
axisymmetric solution that departs from strictly one-dimensional flow near
the conical corners. The inlet total conditions are reproduced to a relative
error of $2.5\times10^{-8}$ in $p_0$ and $7.2\times10^{-9}$ in $T_0$. Over
the final 2\,ms, the maximum steady mass-flow mismatch is $0.055\%$, the
largest monitor drift is $0.081\%$, the largest relative field change is
$0.111\%$, and the maximum transient continuity error is
$6.4\times10^{-8}\%$.

\begin{table}[!htbp]
\centering
\caption{Canonical nozzle: CFD versus isentropic quasi-one-dimensional
theory at $t=6$\,ms. Exit values are area averages on the outlet patch;
throat and inlet values are slab averages. Mass flow is scaled from the
wedge to the full nozzle.}
\label{tab:nozzle}
\small
\begin{tabular}{@{}lrrrr@{}}
\toprule
Quantity & CFD & Quasi-1-D & Difference & Registered limit \\
\midrule
Inlet Mach number$^{a}$ & 0.2528 & 0.2558 & $-1.2\%$ & -- \\
Throat Mach number & 1.029 & 1 & -- & $0.9<M_t<1.1$ \\
Exit Mach number & 1.5133 & 1.5044 & $+0.59\%$ & $<3\%$ \\
Exit static pressure (kPa) & 53.14 & 54.13 & $-1.83\%$ & $<5\%$ \\
Exit static temperature (K) & 205.76 & 206.52 & $-0.37\%$ & $<3\%$ \\
Exit velocity (m\,s$^{-1}$) & 435.09 & 433.36 & $+0.40\%$ & $<3\%$ \\
Mass flow (kg\,s$^{-1}$) & 1.541 & 1.558 & $-1.11\%$ & $<3\%$ \\
\bottomrule
\multicolumn{5}{@{}l}{\footnotesize $^{a}$Not a registered check; theory value computed from the inlet area ratio.}
\end{tabular}
\end{table}

\begin{figure}[!htbp]
    \centering
    \includegraphics[width=0.78\linewidth]{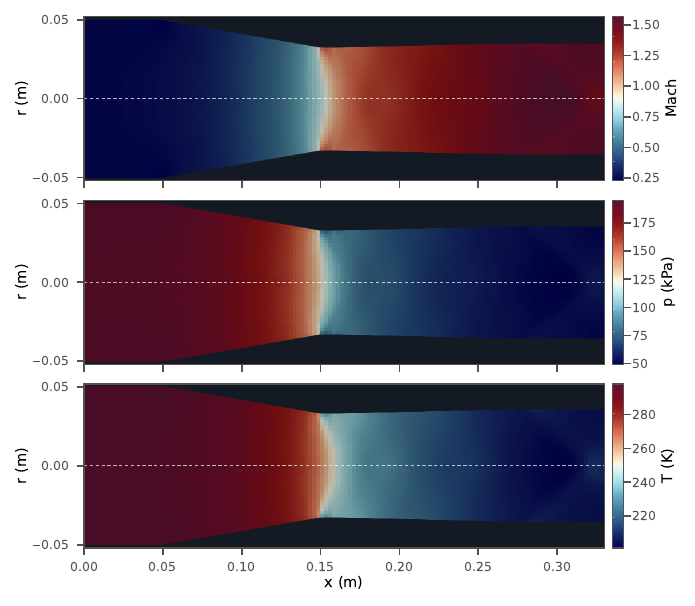}
    \caption{Canonical agent nozzle run at $t=6$\,ms: Mach number, static
    pressure, and static temperature from the archived ParaView renderings of
    the native fields (fixed color ranges recorded in the field-evolution
    manifest), with new axes, from the recorded camera view. The axisymmetric wedge
    solution is mirrored about the axis (dashed). The sonic line sits at the
    throat entrance and the diverging section is supersonic throughout.}
    \label{fig:nozzle_fields}
\end{figure}

A separate archived run of the same specification illustrates the
corrective loop (Fig.~\ref{fig:nozzle}c). Executed initially only to
$t=1$\,ms, it passed 17 of 20 checks but failed the steady mass-balance
($1.35\%$ against $0.1\%$), monitor-stationarity ($2.0\%$ against $0.2\%$),
and field-stationarity ($1.2\%$ against $0.2\%$) criteria. At this
endpoint the exit-state differences from quasi-one-dimensional theory were
already below $0.7\%$. Agreement with the reference therefore did not
establish that the flow had settled, and the stationarity evidence
determined the outcome. After an approved continuation to 6\,ms, all three
metrics fell below their limits (ratios 0.59, 0.41, and 0.66) and the run
was accepted. Figure~\ref{fig:nozzle_fields} shows the accepted canonical
fields.

\subsection{Mach-2 Forward-Facing Step}
\label{sec:step_results}

\begin{figure}[!htbp]
    \centering
    \includegraphics[width=\linewidth]{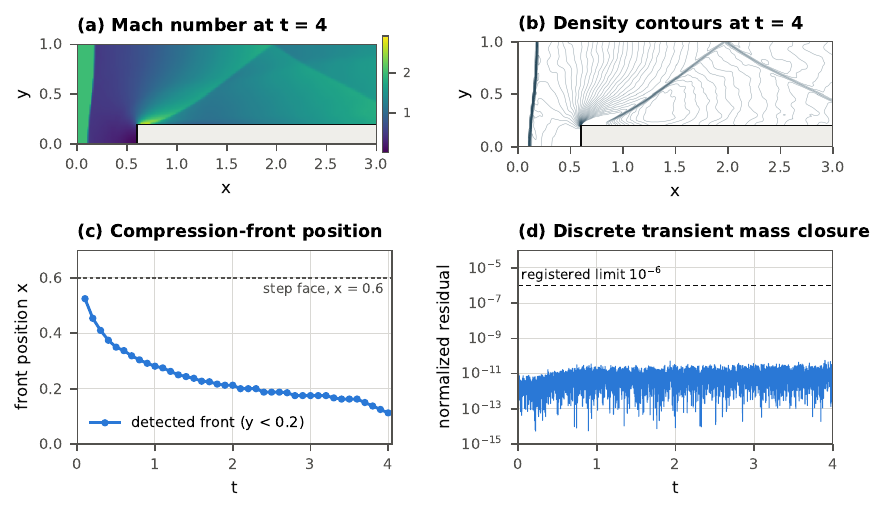}
    \caption{Mach-2 forward-facing step. (a) Mach number and (b) density
    contours at $t=4$, reproduced from the archived pipeline renderings of the
    native fields with new axes. (c) Position of the detected compression
    front below the step height. (d) Normalized discrete closure residual of
    Eq.~(\ref{eq:transient_mass}) at every time step, with the registered
    limit.}
    \label{fig:step}
\end{figure}

The canonical step run completed to $t=4$ in 5,428 steps and 102\,s of wall
time, passing all 17 hard checks and the three compression checks
(\texttt{PASS\_2D\_FORWARD\_STEP}, deterministic \texttt{ACCEPT}). The
realized inlet Mach number was 2.000006, and the maximum Courant number was
0.202 against the 0.2 controller target. Figure~\ref{fig:step} shows the
flow at the registered horizon. A detached compression front spans the
channel ahead of the step, an expansion fan forms at the step shoulder, and
an oblique shock from the shoulder region reflects from the upper symmetry
plane near $x\approx1.95$. Across the detected front, pressure, density,
and temperature rise by factors of 5.18, 2.87, and 1.81. These values exceed
the stationary normal-shock ratios for Mach 2 (4.50, 2.67, and 1.69); the
difference is qualitatively consistent with a front that is moving
upstream, but these gradient-based regional measurements are not
grid-converged and are not used as acceptance bands.

The detected front moves upstream monotonically, from $x=0.525$ at $t=0.1$
to $x=0.113$ at $t=4$ (Fig.~\ref{fig:step}c; snapshots in Fig.~\ref{fig:step_evolution}, Appendix~\ref{app:setup}).
A quasi-one-dimensional
estimate explains this behavior. Behind a normal shock at Mach 2 the flow
has $M=0.577$, whose sonic area is $0.822$ of the channel height. This
exceeds the $0.8$ opening above the step, so the subsonic post-shock flow
cannot pass the contraction and the front is displaced upstream. At Mach 3
the corresponding sonic area is $0.719$, which is why a standing detached
shock is possible in the classical Mach-3 configuration. The canonical
Mach-2 solution is therefore genuinely transient at $t=4$. The registered
contract evaluates health, conservation, and compression structure at that
horizon and applies no stationarity criterion to this family.

Mass conservation closes to round-off in the storage-aware form of
Eq.~(\ref{eq:transient_mass}). The maximum normalized closure residual is
$5.7\times10^{-11}$ (99th percentile $3.0\times10^{-11}$), against the
registered $10^{-6}$ limit (Fig.~\ref{fig:step}d). The cumulative defect is
$8\times10^{-14}$ of the initial mass, and the largest flux through the
impermeable boundaries is $1.1\times10^{-17}$. At $t=4$ the instantaneous
outlet mass flux is $16\%$ below the inlet flux, a difference balanced by
mass accumulating behind the advancing front. A test based on inlet--outlet
equality would have rejected a correctly conservative transient solution.
The closure is a discrete conservation identity
(Appendix~\ref{sec:step_setup}): it confirms that mass is fully accounted
for, not that the solution is accurate.

\subsection{Turbulent Cube Diagnostic Study}
\label{sec:cube_results}

\begin{figure}[!htbp]
    \centering
    \includegraphics[width=\linewidth]{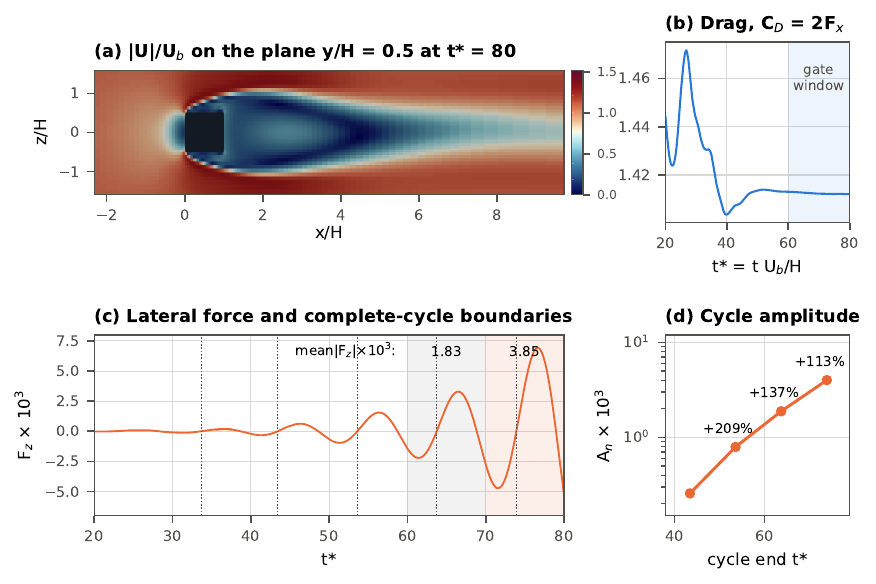}
    \caption{Turbulent cube diagnostic study. (a) Velocity magnitude on the
    plane $y/H=0.5$ at $t^*=80$, from the archived ParaView rendering of the
    native field with new axes. (b) Drag coefficient history; the shaded band
    is the retrospectively registered 20-unit window. (c) Lateral force, with the two
    half-windows of the retrospectively registered growth test shaded and the mean $|F_z|$ of
    each labeled; dotted lines mark the upward zero crossings that bound
    complete cycles. (d) Complete-cycle amplitudes from
    Eq.~(\ref{eq:cycle_amplitude}) and their successive changes.}
    \label{fig:cube}
\end{figure}

The cube calculation completed its bounded extension to $t^*=80$ without
numerical failure. In the final two segments the maximum Courant number was
0.79, the largest reported continuity error was $2\times10^{-12}$, and the
outlet and inlet volume fluxes balanced to a relative $3\times10^{-11}$.
Instantaneous fields at $t^*=80$ show a physically plausible horseshoe
region, separated shear layers from the leading edges, and a recirculating
wake (Fig.~\ref{fig:cube}a). The streamwise force also appears settled. Over
the retrospectively registered window $t^*\in[59.98, 79.98]$, the mean drag coefficient is
$C_D=F_x/(\tfrac12\rho U_b^2H^2)=2F_x=1.412$ and its drift is $0.075\%$ of the mean, well within the
$2\%$ limit, and the mean lateral force is $0.063\%$ of the drag (limit
$5\%$).

The lateral force shows that the flow has not developed
(Fig.~\ref{fig:cube}c). $F_z$ oscillates with an almost constant period of
about 10.15 time units ($St\approx0.098$), while its envelope grows: the mean
of $|F_z|$ rises from $1.83\times10^{-3}$ in the first half of the window
to $3.85\times10^{-3}$ in the second, a ratio of 2.10 against the registered
limit of 1.25. The gate therefore returns \texttt{STILL\_DEVELOPING}, and the
run is rejected. The supplementary complete-cycle audit agrees
(Fig.~\ref{fig:cube}d). Successive complete-cycle amplitudes grow by
$209\%$, $137\%$, and $113\%$, and the last comparable pair exceeds the
$10\%$ audit limit by an order of magnitude. The growth rate is declining
but the mode has not saturated. The two diagnostics are different
measurements and are reported separately. Only the half-window ratio is part
of the registered gate.

The result illustrates why the assessment cannot rely on a single signal.
Solver completion, small continuity errors, a stable mean drag, and plausible
instantaneous contours all coexist here with a lateral mode whose amplitude
more than doubles every cycle. Block averages hide it as well: between the
$t^*=60$--70 and 70--80 averages the mean drag changes by only $0.04\%$ and
wake-probe means by at most $0.002\,U_b$, while the root-mean-square lateral
force increases by $110\%$. Instantaneous lateral-velocity fields on the plane
$y/H=0.5$ at $t^*=20$, 40, 60, and 80 are almost indistinguishable on a common
color range (only $t^*=80$ is shown, in Fig.~\ref{fig:flowpilot_architecture}c), so contours alone would not reveal the growing mode
that the force history exposes.

\paragraph{Agent diagnosis of the cube evidence.} Because the cube campaign
ran outside the agent loop, we applied a diagnosis call written for this
replay, in the style of the agent's diagnosis stage, to its archived evidence
afterwards, without running any CFD. Its system instruction states that solver
completion does not prove a developed flow and asks the model to consider every
force component, not only the drag. The evidence packet
contained solver health from the seven run segments (all ended normally;
maximum Courant number 0.80; maximum global continuity error
$5.4\times10^{-11}$), force statistics in six 10-unit blocks, and the
half-cycle extrema of $F_z$, but no gate verdict and no threshold. The model
was called three times with the same packet at temperature 0
(Table~\ref{tab:cube_llm}). Twice it diagnosed \texttt{STILL\_DEVELOPING} and
proposed \texttt{CONTINUE\_RUN}, noting that $F_x$ and $F_y$ had settled while
the $F_z$ extrema grew in every half-cycle; this agrees with the registered
gate, which the model never saw, although its instruction pointed to all force
components. Once it read the same growth as a numerical
instability, diagnosed \texttt{NUMERICALLY\_UNHEALTHY}, and proposed
\texttt{FAIL\_SAFELY}, although the solver-health evidence was clean; the six
warning messages counted in the packet are routine OpenFOAM time-precision
notices, but the packet reported only their number. No call proposed
\texttt{ACCEPT}, so the validator, which restricts \texttt{ACCEPT} but permits
stopping or continuing a run, approved all three, and no action was executed
in this replay. Each call took about 2\,s and used about 4,800 tokens.
Temperature 0 therefore did not make the output repeatable: identical inputs
produced two different diagnoses, one of which misread clean solver health as
instability. Acceptance is decided by the registered gate, not by these
diagnoses.

\begin{table}[!htbp]
\centering
\caption{LLM diagnosis of the archived cube evidence: all three calls with
an identical packet (temperature 0). The registered gate returns
\texttt{STILL\_DEVELOPING} (lateral-force growth ratio 2.10 against 1.25).
No action was executed.}
\label{tab:cube_llm}
\footnotesize
\begin{tabular}{@{}clllrr@{}}
\toprule
Call & Diagnosis & Proposed action & Validator & Latency (s) & Tokens \\
\midrule
1 & \texttt{STILL\_DEVELOPING} & \texttt{CONTINUE\_RUN} & Approved & 2.5 & 4,805 \\
2 & \texttt{NUMERICALLY\_UNHEALTHY} & \texttt{FAIL\_SAFELY} & Approved & 2.2 & 4,709 \\
3 & \texttt{STILL\_DEVELOPING} & \texttt{CONTINUE\_RUN} & Approved & 1.7 & 4,764 \\
\bottomrule
\end{tabular}
\end{table}

\section{Limitations}
\label{sec:limitations}

The demonstration covers three CFD studies. Nozzle and forward-facing-step
flows are the routable families; their accepted runs pass the registered
checks, which is narrower than physical validation. The cube is supporting diagnostic
evidence and is not routable. Each new physics family requires its own
scientific contract, and acceptance under a family contract is not a general
statement of CFD correctness. The accepted runs are single-mesh
verifications. The nozzle family rests on a frozen four-mesh reference
campaign, but no grid-convergence index is reported, and the step family has
neither a registered cross-grid criterion nor a quantitative reference-field
comparison. The
cube wall treatment is uncertain: at $t^*=80$ the cube-surface $y^+$ has a
median of 24.7 and a range of 2.8--96.5, with $67\%$ of the area below
$y^+=30$. Its 20-unit gate window contains only about two periods of the lateral
oscillation, enough to show continued growth but too short for a statistical
stationarity claim. A provisional comparison of its evolving mean wake with
ERCOFTAC centerline profiles is not satisfactory, so no experimental
validation is claimed. The model's diagnoses are not guaranteed to be repeatable: in the cube
diagnosis (Sec.~\ref{sec:cube_results}) one of three identical calls gave a different diagnosis, whereas all five
arm-B repeats on the same cube evidence agreed in the controller comparison, and the proposed corrective
action varied across repeats at five of 18 planted faults
(Sec.~\ref{sec:controller}). General CAD/STEP-to-CFD execution is not
demonstrated. All archived agent sessions are development cases, run while
the corresponding pipeline was being built, and the cube gate was registered
after the cube data existed (Appendix~\ref{app:repro}); no request was held
out. Finally, the controller comparison (Sec.~\ref{sec:controller}) uses one
model, replays archived states rather than new runs, and plants faults in one
family. In the check-informed B/B$'$ configuration the model never proposed
acceptance contrary to a failed check, so the enforcement layer was not
tested against a model that proposes an unjustified acceptance. Nor does
the comparison establish an incremental benefit of the model over a fixed
rule (Sec.~\ref{sec:controller}), its two configurations also differ in
prompting, and the corrections the model proposed for planted faults were not
executed. In the agent sessions, one repair call (N4) was not recorded, and one
accepted step run (S7) required supervised resumption after a mid-campaign
validator fix. Comparisons across models, on held-out
requests, and with executed corrections remain future work.

\section{Conclusion}
\label{sec:conclusion}

The aim of this work is CFD that can be automated from request to report and
still be trusted. CFD Forge links engineering requests to OpenFOAM case construction,
execution, diagnosis, correction, and reporting, while keeping scientific
acceptance in deterministic, family-specific contracts. Language-model
reasoning interprets requests and evidence and proposes bounded actions;
validators decide whether those actions are admissible and whether a result
is accepted. Under this division, a converging--diverging nozzle was accepted
with exit-state and mass-flow differences from quasi-one-dimensional theory
below $2\%$ after its
stationarity evidence, not its reference agreement, determined when the run
was complete. A transient Mach-2 forward-facing-step flow was accepted with
storage-aware mass closure at round-off. Within both families, the agent also built and
assessed new cases from text requests that changed the geometry and the
operating conditions, and their results followed the expected physical
trends. For a three-dimensional turbulent cube run outside the agent loop, a
retrospectively registered stationarity gate withheld acceptance because a
growing lateral mode showed that solver completion and a settled mean drag
did not establish a developed flow. In a controller comparison of two configurations of the same model, a
model-only configuration given a filtered evidence packet accepted most of the
planted faults, whereas the CFD Forge configuration, with the check results
and the validators, accepted none; because the configurations also differ in
prompting, this does not isolate the effect of the check results. A fixed rule
reached the same verdicts as CFD Forge, so these results do not establish an
incremental benefit of the model in acceptance decisions. Case files,
recorded model calls, action rulings, and field-evolution records keep each
decision auditable. These elements provide a basis for high-throughput simulation,
active learning, and design-optimization loops in which automated
evaluations must remain scientifically accountable.

\appendix
\section{Detailed Simulation Setup and Acceptance Criteria}
\label{app:setup}

\begin{table}[!htbp]
\centering
\caption{Simulation setup. All cases use OpenFOAM Foundation v14 through
\texttt{foamRun}. The scheme, solution, thermophysical and mesh files quoted
here match the SHA-256 digests recorded for the native cases in the
field-evolution manifests; cube boundary types are taken from the preserved
initial-condition audit of its predecessor pilot.}
\label{tab:setup}
\small
\setlength{\tabcolsep}{3.5pt}
\begin{tabularx}{\linewidth}{@{}>{\raggedright\arraybackslash}p{2.05cm}
  >{\raggedright\arraybackslash}X >{\raggedright\arraybackslash}X
  >{\raggedright\arraybackslash}X@{}}
\toprule
 & Nozzle & Forward-facing step & Surface-mounted cube \\
\midrule
Model & Inviscid compressible Euler, perfect gas ($\gamma=1.4$, $R=287$\,J\,kg$^{-1}$\,K$^{-1}$); \texttt{shockFluid}
 & Inviscid compressible Euler, normalized perfect gas ($\gamma=1.4$, sound speed 1); \texttt{shockFluid}
 & Incompressible unsteady Reynolds-averaged Navier--Stokes (URANS), $k$--$\omega$ shear-stress transport (SST)~\cite{menter1994sst}, $\nu=2.5\times10^{-5}$, $Re_H=40{,}000$; \texttt{incompressibleFluid} \\
Mesh & 5$^\circ$ wedge, $132\times16=2{,}112$ cells (1,980 hexahedra and 132 axis-adjacent prisms); max.\ aspect ratio 2.46, non-orthogonality $9.6^\circ$, skewness 0.33
 & Three blocks, uniform $\Delta x=\Delta y=0.0125$, 16,128 hexahedra, one \texttt{empty} spanwise layer
 & Rectilinear, $116\times40\times124$ minus the $20^3$ cube cells $=567{,}360$ hexahedra; max.\ aspect ratio 18.4, six partitions \\
Inlet & \texttt{totalPressure} $p_0=200$\,kPa, \texttt{totalTemperature} $T_0=300$\,K, axial-direction velocity
 & Fixed $p=1$, $T=1$, $\mathbf{U}=(2,0,0)$
 & Developed $\mathbf{U}$, $k$, $\omega$ mapped from a precursor channel; $\partial p/\partial n=0$ \\
Outlet & Zero gradient for $p$, $T$, $\mathbf{U}$ (30\,kPa ambient not imposed)
 & Zero-gradient $p$; \texttt{inletOutlet} $\mathbf{U}$, $T$
 & Fixed gauge $p$; zero gradient for $\mathbf{U}$, $k$, $\omega$ \\
Other boundaries & Slip wall; \texttt{wedge} front/back; axis
 & Step: slip; top/bottom: \texttt{symmetryPlane}
 & Floor, roof ($y/H=2$), side walls ($z/H=\pm12$), cube: no-slip with \texttt{kqR}/\texttt{omega}/\texttt{nutUSpalding} wall functions \\
Numerics & Kurganov flux~\cite{kurganov2000central}, Minmod reconstruction, Euler time step, $\mathrm{Co}_{\max}=0.4$
 & Kurganov flux, van Leer ($\rho$, $T$) and van Leer-V ($\mathbf{U}$) reconstruction, Euler, $\mathrm{Co}_{\max}=0.2$
 & Backward time step, PIMPLE (2 outer, 2 pressure correctors), linear-upwind $\mathbf{U}$, upwind $k,\omega$, $\mathrm{Co}_{\max}=0.8$, $\Delta t\le0.02$ \\
Horizon & $t=6$\,ms (30,471 steps) & $t=4$ (5,428 steps) & $t^*=80$ \\
Assessment basis & Quasi-1-D analytical consistency and registered numerical checks
 & Registered numerical checks and directional compression diagnostics
 & Registered stationarity/development gate \\
\bottomrule
\end{tabularx}
\end{table}

\begin{figure}[!htbp]
    \centering
    \includegraphics[width=0.88\linewidth]{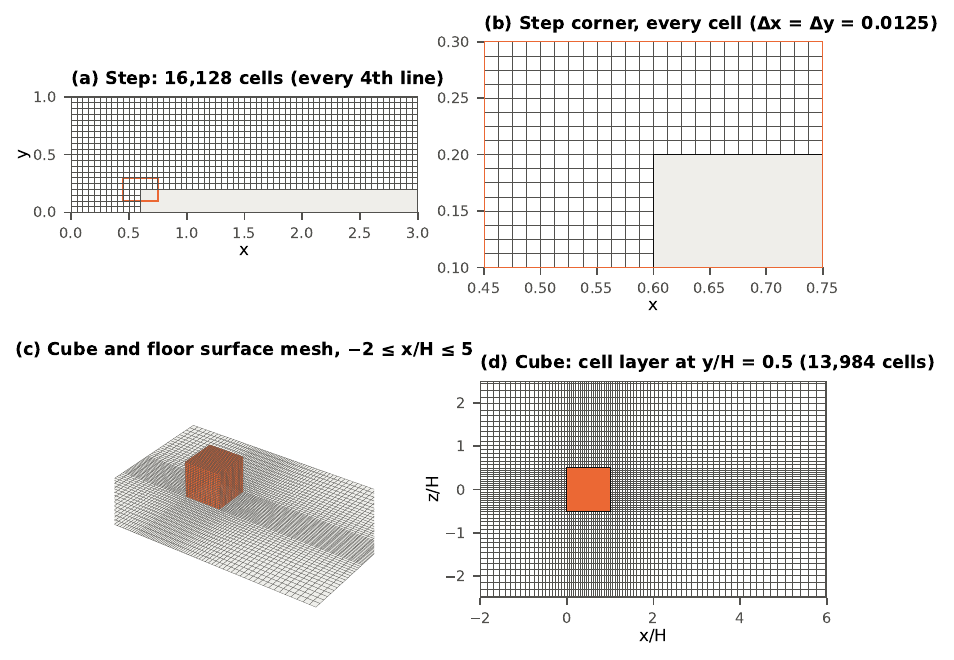}
    \caption{Step and cube meshes, plotted from the native OpenFOAM
    \texttt{polyMesh} files. (a) Forward-facing-step domain (every fourth grid
    line) with the step in gray and (b) every cell near the step corner; the
    mesh topology files are identical to those of the agent's Mach-2 case.
    (c) Cube and floor boundary faces near the cube and (d) the cell layer
    starting at $y/H=0.5$, showing the clustering around the cube and in the
    near wake.}
    \label{fig:meshes_step_cube}
\end{figure}

\subsection{Converging--Diverging Nozzle}
\label{sec:nozzle_setup}

The nozzle has inlet, throat, and exit radii of 0.0500, 0.0326, and
0.0354\,m over a length of 0.33\,m ($A_e/A^*=1.179$, nominal $p_0/p_a=6.67$);
the remaining setup is listed in Table~\ref{tab:setup}, and the wedge mesh is
shown in Fig.~\ref{fig:case_geometries}a. The 30\,kPa ambient pressure is used
only in the operating-regime assessment; it is not imposed at the outlet, and
the external jet is not simulated. Quasi-one-dimensional theory supplies the
initial fields and the analytical comparison but does not constrain the exit
state.

The twenty registered checks cover mesh validity, solver completion, positive
and finite thermodynamic fields, inlet total conditions, a sonic throat
($0.9<M_t<1.1$), a supersonic outlet (minimum face-normal Mach number above
1.05), stagnation enthalpy, and flux consistency. Over the final 2\,ms they
require a steady mass-flow mismatch below $0.1\%$, a normalized transient
continuity error below $0.01\%$, and monitor drift and relative field changes
below $0.2\%$; the exit-pressure difference from theory must stay below $5\%$
and the other theory comparisons below $3\%$. A passing run is
\texttt{PASS\_SINGLE\_MESH}, not a grid-convergence study. The recipe was
frozen beforehand from an LLM-free reference campaign on four refined wedge
meshes and three controls (halved Courant number, halved wedge angle, and a
$2\%$ startup perturbation), which required last-pair changes below $0.5\%$
and control changes below $0.2\%$ (Table~\ref{tab:nozzle_grid}). The theory
tolerances are screening bounds for a finite-angle conical flow.

\begin{table}[!htbp]
\centering
\caption{Frozen LLM-free nozzle reference campaign: mesh refinement at
$t=6$\,ms. Mesh 1 is identical to the agent's canonical run. Coarse-mesh
controls (halved Courant number, halved wedge angle, $2\%$ startup
perturbation) changed all reported quantities by at most $0.036\%$.}
\label{tab:nozzle_grid}
\small
\begin{tabular}{@{}lrrrrr@{}}
\toprule
Mesh (axial $\times$ radial) & Cells & Throat $M$ & Exit $M$ & Exit $p$ (kPa) & $\dot m$ (kg\,s$^{-1}$) \\
\midrule
$132\times16$ & 2,112 & 1.0290 & 1.5133 & 53.14 & 1.5409 \\
$198\times24$ & 4,752 & 1.0432 & 1.5189 & 52.78 & 1.5378 \\
$264\times32$ & 8,448 & 1.0512 & 1.5221 & 52.57 & 1.5361 \\
$330\times40$ & 13,200 & 1.0555 & 1.5244 & 52.43 & 1.5349 \\
\midrule
Last-pair change & & $0.41\%$ & $0.15\%$ & $0.27\%$ & $0.07\%$ \\
\bottomrule
\end{tabular}
\end{table}

\subsection{Mach-2 Forward-Facing Step}
\label{sec:step_setup}

The channel has length 3 and height 1, with a step of height 0.2 at $x=0.6$,
meshed with 16,128 cells and one \texttt{empty} spanwise layer
(Fig.~\ref{fig:meshes_step_cube}a,b; Table~\ref{tab:setup}). The normalized
gas has unit sound speed, so the inlet state $p=T=1$, $\mathbf{U}=(2,0,0)$ is
Mach 2. The numerical recipe is copied unchanged from the Foundation v14
\texttt{shockFluid/forwardStep} tutorial and hash-checked at run time; it was
frozen after a manual reproduction of the Mach-3 tutorial on the same mesh,
whose shock structure is qualitatively consistent with the classical
problem~\cite{woodward1984strong} but is not a quantitative reference. The
case differs from that Mach-3 problem and is not presented as a reproduction
of it.

The checks cover planar mesh topology, the initial state, realized supersonic
inflow, field positivity, solver health, wall impermeability, completion of the
horizon ($t=4$), and measurable increases of pressure, density, and
temperature across the detected front. Conservation is tested with the
transient balance
\begin{equation}
    \frac{\mathrm{d}}{\mathrm{d}t}
    \int_{\Omega}\rho\,\mathrm{d}V
    +
    \int_{\partial\Omega}
    \rho\mathbf{U}\cdot\mathbf{n}\,\mathrm{d}S
    = 0,
    \label{eq:transient_mass}
\end{equation}
evaluated from two records written at every time step: the domain mass
$M^n=\sum_i\rho_i^nV_i$ and the net boundary mass flux $\Phi^n$ summed from
the face fluxes. The residual $r^n=(M^n-M^{n-1})/\Delta t^n+\Phi^n$, normalized
by the larger of the inlet and outlet fluxes, must stay below $10^{-6}$;
sample pairs that straddle a restart are never differenced. Because the solver
advances a conservative finite-volume update with explicit Euler time
integration, this identity holds to round-off when every term is included, so
it detects accounting errors---a missing boundary flux, a spurious source, or
a broken restart---but not discretization error. Unequal instantaneous inlet
and outlet fluxes are therefore not a conservation failure when they are
balanced by a change in stored mass. The observed residuals are near
$10^{-11}$. The compression checks are directional, not a quantitative
reference-field comparison.

\begin{figure}[!htbp]
    \centering
    \includegraphics[width=0.92\linewidth]{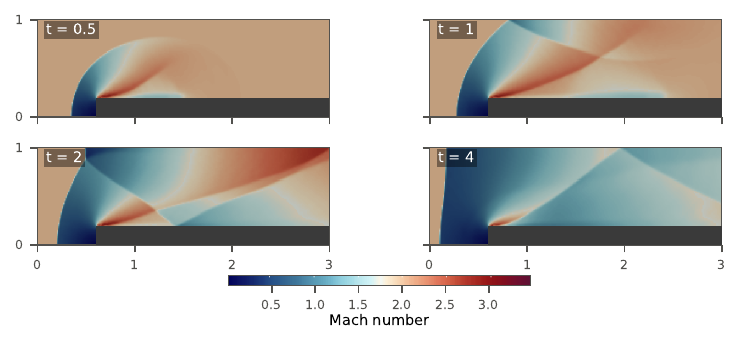}
    \caption{Evolution of the agent's Mach-2 step solution: Mach number at
    $t=0.5$, 1, 2, and 4 from the archived ParaView renderings of the native
    fields, on one fixed color range. The detached front moves upstream
    throughout, while the shoulder expansion and the reflected oblique shock
    reorganize behind it.}
    \label{fig:step_evolution}
\end{figure}

\subsection{Turbulent Cube Diagnostic Study}
\label{sec:cube_setup}

The cube ($H=1$) sits in a channel of height $2H$ spanning $x/H\in[-6,16]$
and $z/H\in[-12,12]$, at $Re_H=U_bH/\nu=40{,}000$, comparable to the
configuration of Martinuzzi and Tropea~\cite{martinuzzi1993cube}. The mesh has
567,360 cells (Fig.~\ref{fig:meshes_step_cube}c,d), and the inflow is mapped
from a precursor channel (Table~\ref{tab:setup}). The campaign continues a
predecessor pilot to $t^*=tU_b/H=80$, and force histories are analyzed from
$t^*=20$.

Forces are the pressure-plus-viscous forces on the cube in units with
$\rho=U_b=H=1$, so $C_D=2F_x$, and $F_z$ is plotted in these units. The stationarity gate, registered
retrospectively (Appendix~\ref{app:repro}), is applied over the final 20 time
units: the least-squares drift of each force component must stay below $2\%$
of the mean streamwise force, the mean lateral force below $5\%$ of it, and the
mean of $|F_z|$ over the second half of the window must not exceed $1.25$
times its first-half value. A supplementary audit, which is not part of the
registered gate, compares complete lateral-force cycles between successive
upward zero crossings of $F_z$, with amplitude
\begin{equation}
    A_n =
    \frac{F_{z,\max,n}-F_{z,\min,n}}{2},
    \label{eq:cycle_amplitude}
\end{equation}
and requires successive comparable amplitudes to change by less than $10\%$;
incomplete cycles are excluded. Mean drag, six wake probes, block-mean fields,
and fluctuation measures provide complementary evidence.
The cube is a diagnostic stress test, not an accepted turbulent reference
solution.

\section{Model Records}
\label{app:records}

Figures~\ref{fig:walk_spec} and~\ref{fig:trace} reproduce two archived model
outputs referred to in the main text, and Table~\ref{tab:timeline} gives the
timeline of the walkthrough session.

\begin{figure}[!h]
\centering
\begin{minipage}{0.96\linewidth}
\footnotesize
\begin{verbatim}
model output (gemini-3.5-flash-lite, schema-constrained, temperature 0)
  case_name: mach2p0_step0p20     mach: 2.0   step_height: 0.2   step_x: 0.6
  length, height, pressure, temperature, end_time, max_co, nx, ny: null
  requests_unsupported_physics: false
  reporting_objectives:
    - inspect the transient shock/compression structure
    - determine whether the final result passes deterministic scientific validation
resolved specification
  length 3.0  height 1.0  nx 240  ny 80  pressure 1.0  temperature 1.0
  end_time 4.0  max_co 0.2  write_interval 0.1  physics: inviscid_euler
field provenance
  stated_by_user:                mach, step_height, step_x
  inferred_from_family_defaults: end_time, height, length, max_co, nx, ny,
                                 pressure, temperature, write_interval
\end{verbatim}
\end{minipage}
\caption{Request interpretation in the walkthrough session (abridged from the
archived \texttt{llm\_interpretation.json}). The model returns only what the
request states; family defaults complete the specification, and the origin of
every field is recorded.}
\label{fig:walk_spec}
\end{figure}

\begin{figure}[!h]
\centering
\begin{minipage}{0.96\linewidth}
\footnotesize
\begin{verbatim}
diagnosis: UNCONVERGED            confidence: high
evidence_used:
  - mass imbalance is 1.35 percent, exceeding the steady balance criterion
  - passes_stationarity is false
  - mass flow range and drift indicate ongoing evolution
  - no local mesh underresolution flagged by baseline metrics
competing_hypotheses:
  - insufficient_run_time
      for:     monitors and fields are not yet stationary;
               mass imbalance and drift are slowly decreasing
  - local_mesh_underresolution
      for:     strong gradients present near throat
      against: checkMesh passed cleanly; no candidate underresolved
               regions flagged by feedback profile diagnostics
action: CONTINUE_RUN
action validator: APPROVED ("permitted for a healthy evolving solution")
scientific validator: FAIL (steady_mass_balance, monitors_stationary,
                            fields_stationary)
\end{verbatim}
\end{minipage}
\caption{Archived \texttt{gemini-3.5-flash-lite} diagnosis record for the
nozzle continuation run at $t=1$\,ms (abridged and reformatted from
\texttt{agent\_decision.json}). The model separates
insufficient run time from local under-resolution using the evidence packet,
and its proposal is executed only after the action validator approves it.}
\label{fig:trace}
\end{figure}

\begin{table}[!h]
\centering
\caption{Timeline of the walkthrough session, from the archived event log.
``LLM'' marks a model call (the image observer is a multimodal call whose
model identifier was not recorded; provider usage records indicate
\texttt{gemini-3.6-flash} for the step sessions).}
\label{tab:timeline}
\small
\begin{tabular}{@{}rlll@{}}
\toprule
Time (s) & Stage & Performed by & Outcome \\
\midrule
0.0 & Request received & -- & 313-character prompt \\
5.9 & Request interpretation & LLM & Mach 2, $h=0.2$, $x=0.6$ \\
5.9 & Scope gate & Deterministic & 14/14 checks, in scope \\
22.0 & Case and mesh generation & Deterministic & 16,128 cells, recipe hash-checked \\
28.3 & Mesh check & OpenFOAM & Mesh OK \\
134.2 & Solver run & OpenFOAM & Completed to $t=4$ \\
148.9 & Evidence extraction & Deterministic & 17/17 hard checks \\
158.8 & Field-image observation & LLM (multimodal) & Observation complete \\
161.7 & Diagnosis and action & LLM & \texttt{HEALTHY\_COMPLETE} $\rightarrow$ \texttt{ACCEPT} \\
161.7 & Action validation & Deterministic & Approved \\
161.7 & Scientific verdict & Deterministic & \texttt{PASS\_2D\_FORWARD\_STEP} \\
165.4 & Engineering summary & LLM & Report written \\
\bottomrule
\end{tabular}
\end{table}

\section{Reproducibility, Session Ledger, and Development History}
\label{app:repro}

\paragraph{Models and calls.} All recorded text calls used
\texttt{gemini-3.5-flash-lite} through the Google Gemini API
(\texttt{google-genai} SDK) at temperature 0, with a Pydantic response schema
for every stage except the plain-text summary. The field observer's model is
inferred from the provider's usage records (Sec.~\ref{sec:model}).
The nozzle sessions ran on 21 September 2026, the step sessions on 22
September, the cube calculation on 23--24 September, and the cube diagnosis
(Sec.~\ref{sec:cube_results}) and controller comparison
(Sec.~\ref{sec:controller}) on 1 October 2026. System instructions and schemas
are in \texttt{src/agents/},
\nolinkurl{src/reasoning/forward_step_diagnosis.py}, and
\nolinkurl{paper/cfd_forge/scripts/cube_llm_diagnosis.py}. Each run allowed at
most four iterations of solve, diagnosis, and action. Token counts were
recorded for the cube diagnosis and the controller comparison only.

{\sloppy\paragraph{Action vocabularies.} The nozzle family
admits \texttt{ACCEPT}, \texttt{CONTINUE\_RUN},
\texttt{REQUEST\_DIAGNOSTIC}, \texttt{REFINE\_THROAT},
\texttt{REFINE\_GRADIENT\_REGION}, \texttt{REPAIR\_MESH},
\texttt{RESTART\_CLEAN}, and \texttt{REJECT\_OUTSIDE\_DOMAIN}. The step
family admits \texttt{ACCEPT}, \texttt{CONTINUE\_RUN},
\texttt{EXTEND\_END\_TIME}, \texttt{REDUCE\_MAX\_CO},
\texttt{REFINE\_MESH}, \texttt{REBUILD\_FROM\_VALIDATED\_SPEC},
\texttt{REQUEST\_CLARIFICATION}, \texttt{REJECT\_UNSUPPORTED}, and
\texttt{FAIL\_SAFELY}. \texttt{REFINE\_MESH} is refused unless the current solution is
healthy.\par}

\paragraph{Counting.} Model calls were counted from every call record in the
session archive---the per-session call logs and the per-iteration decision,
mesh-review, and report records---and deduplicated within each run by stage
and latency, because records are copied between session and iteration
folders. Field-observation calls were counted from the per-iteration
observation records. An earlier count taken from the event logs alone (82
calls, 23 proposals) missed the iterations of the nozzle feedback loop and
is superseded by the numbers in Sec.~\ref{sec:agent_results}.

\paragraph{What is reproducible.} The archive reproduces the analyses,
figures, and tables from the recorded evidence, and the native case files
(initial or restart state, mesh, settings, and logs) allow each CFD run to be
repeated with OpenFOAM Foundation v14. The historical agent sessions ran from uncommitted
development trees, so the release does not reproduce the historical agent
execution byte for byte. Model outputs are not reproducible either:
temperature 0 does not make the provider deterministic, as the cube
diagnosis shows (Sec.~\ref{sec:cube_results}). Reproducibility here means
reproducibility of the artifacts and the workflow, not of the exact model
outputs.

\paragraph{Hardware and software.} All calculations ran on one Windows laptop
with an Intel Core i5-11400H processor (six cores), under OpenFOAM Foundation
version 14 (build \texttt{14-7b05503f98a8}) in WSL~2. The nozzle and step runs
were serial; the cube ran on six MPI ranks. The nozzle sessions ran from the
uncommitted development tree between commits \texttt{2d86e7b} and
\texttt{007a71d}, and the step sessions between \texttt{3bd128a} and
\texttt{7f9126a}. The consolidated repository is tagged
\texttt{v0.1-review-baseline}.

\paragraph{Development history and threshold registration.} All archived
agent sessions are development cases, run while the corresponding pipeline was
being built; no request was held out. The nozzle thresholds were committed on
20 September 2026 with the LLM-free reference campaign, before the first
nozzle session, as declared engineering tolerances that may not be relaxed to
pass a run; the nozzle requests were then repeated as the feedback loop was
revised, which is why the reference request appears in four sessions. The
step validator was committed on 22 September 2026, before the first step
session, and changed twice during that campaign: to remove the false-positive
fatal-error check (Sec.~\ref{sec:agent_results}) and to form the mass-closure
residual only within a restart segment. The cube gate
(\texttt{cube-stationarity/1.0.0}) was registered on 28 September 2026, after
the cube calculation had been run. The measured growth ratio is 2.10, so the
gate returns the same verdict for any growth limit below 2.10, including the
registered 1.25; the gate was nonetheless not frozen before the data existed.

\paragraph{Prototype before the registered contracts.} The registered nozzle
family replaced an earlier, less constrained prototype, preserved in the
archive. Given the same nozzle as text, it built the geometry through a
CAD$\rightarrow$Gmsh path, and the model chose the meshing strategy
(Fig.~\ref{fig:case_geometries}b); the resulting 63,747-tetrahedron mesh met
its quality constraint and was accepted. The model's solver setup, however,
was inconsistent for the inviscid request: it selected laminar flow with
Sutherland viscosity while its rationale cited $k$--$\omega$ SST, and it
recorded the 30\,kPa back pressure as the outlet static pressure. After two
continuations the run ended at $t=0.2$\,ms as \texttt{STABLE\_UNCONVERGED}
with a $50.8\%$ mass-flow imbalance. In the registered family the physical
model and boundary conditions are fixed by the contract, and the model
supplies only case parameters. Carrying the Gmsh path through to a passing
solution is left to future work.

\paragraph{Session ledger.} Table~\ref{tab:ledger} lists all 19 sessions.
Proposals are given in order with the action validator's ruling; the final
column gives the archived family status and the resulting decision under
Sec.~\ref{sec:authority}; runs that ended before an approved correction was
carried out are marked ``not accepted''.

\begin{table}[!htbp]
\centering
\caption{Ledger of the archived agent sessions. A = approved, R = refused.
``Repeat'' marks a request repeated during development of the same
pipeline.}
\label{tab:ledger}
\scriptsize
\setlength{\tabcolsep}{2.5pt}
\begin{tabularx}{\linewidth}{@{}>{\raggedright\arraybackslash}p{2.6cm}
  >{\raggedright\arraybackslash}p{3.1cm} >{\raggedright\arraybackslash}X
  >{\raggedright\arraybackslash}p{3.3cm}@{}}
\toprule
Run (sessions) & Request & Proposals (ruling) & Final status $\rightarrow$ decision \\
\midrule
\multicolumn{4}{@{}l}{\emph{Nozzle, 21 September 2026}} \\
N1 \texttt{e2e/A} (1) & Reference nozzle, $p_0=200$\,kPa, 6\,ms & \texttt{ACCEPT} (A) & \texttt{PASS\_SINGLE\_MESH} $\rightarrow$ \texttt{ACCEPT} (canonical run) \\
N2 \texttt{e2e/B} (1) & Exit radius 37\,mm & \texttt{ACCEPT} (A) & \texttt{PASS\_SINGLE\_MESH} $\rightarrow$ \texttt{ACCEPT} \\
N3 \texttt{e2e/C} (1) & $p_0=220$\,kPa & \texttt{ACCEPT} (A) & \texttt{PASS\_SINGLE\_MESH} $\rightarrow$ \texttt{ACCEPT} \\
N4 \texttt{feedback/A} (1) & Repeat of N1, 1\,ms horizon & \texttt{CONTINUE\_RUN} (A); \texttt{REQUEST\_DIAGNOSTIC} (A, not executed) & \texttt{UNEXECUTED\_ACTION\_\allowbreak REQUEST\_\allowbreak DIAGNOSTIC} $\rightarrow$ not accepted \\
N5 \texttt{feedback\_v2/A} (1) & Repeat of N1, 1\,ms horizon & \texttt{CONTINUE\_RUN} (A); session ended before the continuation & \texttt{FAIL} at 1\,ms, continuation not run $\rightarrow$ not accepted \\
N6 \texttt{v2\_hotfix/A} (1) & Repeat of N1, 1\,ms horizon & \texttt{CONTINUE\_RUN} (A); \texttt{CONTINUE\_RUN} (A); \texttt{ACCEPT} (A) at 6\,ms & \texttt{PASS\_SINGLE\_MESH} $\rightarrow$ \texttt{ACCEPT} (continuation run, Sec.~\ref{sec:nozzle_results}) \\
N7 \texttt{feedback\_final/B} (1) & Repeat of N2 & \texttt{ACCEPT} (A); \texttt{ACCEPT} (A) & \texttt{PASS\_SINGLE\_MESH} $\rightarrow$ \texttt{ACCEPT} \\
N8 \texttt{feedback\_final/C} (1) & Repeat of N3 & \texttt{ACCEPT} (A); \texttt{ACCEPT} (A) & \texttt{PASS\_SINGLE\_MESH} $\rightarrow$ \texttt{ACCEPT} \\
\midrule
\multicolumn{4}{@{}l}{\emph{Forward-facing step, 22 September 2026}} \\
S1 \texttt{live\_run\_01} (1) & Mach 2.5, $h=0.15$ & \texttt{FAIL\_SAFELY} (A) & \texttt{FAIL} (false-positive fatal check) $\rightarrow$ \texttt{REJECT}; reanalysis after the fix: \texttt{PASS\_2D\_FORWARD\_STEP} \\
S2 \texttt{case\_B} (1) & Mach 2, $h=0.2$ & \texttt{ACCEPT} (A) & \texttt{PASS\_2D\_FORWARD\_STEP} $\rightarrow$ \texttt{ACCEPT} (walkthrough, Sec.~\ref{sec:walkthrough}) \\
S3 \texttt{case\_C} (1) & Mach 3.5, $h=0.2$ & \texttt{ACCEPT} (A) & \texttt{PASS\_2D\_FORWARD\_STEP} $\rightarrow$ \texttt{ACCEPT} \\
S4 \texttt{case\_E} (1) & Mach 3, $h=0.1$ & \texttt{ACCEPT} (A) & \texttt{PASS\_2D\_FORWARD\_STEP} $\rightarrow$ \texttt{ACCEPT} \\
S5 \texttt{case\_F} (1) & Mach 3, $h=0.3$, $x=0.6$ & \texttt{FAIL\_SAFELY} (A) & \texttt{FAIL} (no measurable front) $\rightarrow$ \texttt{REJECT} \\
S6 \texttt{case\_G} (1) & Mach 3, $h=0.3$, $x=1.0$ & \texttt{ACCEPT} (A) & \texttt{PASS\_2D\_FORWARD\_STEP} $\rightarrow$ \texttt{ACCEPT} \\
S7 \texttt{case\_H} (4, resumed) & Mach 3, $h=0.2$, staged horizon to $t=4$ & \texttt{EXTEND\_END\_TIME} (A); \texttt{CONTINUE\_RUN} (R); \texttt{EXTEND\_END\_TIME} (A); \texttt{FAIL\_SAFELY} (A); \texttt{EXTEND\_END\_TIME} (R, budget); \texttt{EXTEND\_END\_TIME} (A); \texttt{ACCEPT} (A) & \texttt{PASS\_2D\_FORWARD\_STEP} $\rightarrow$ \texttt{ACCEPT} after supervised resumption and a mass-closure fix \\
S8 \texttt{case\_I} (1) & Mach 3, $h=0.2$, sensitivity request & \texttt{REFINE\_MESH} (A); \texttt{ACCEPT} (R) & \texttt{STOPPED\_ACTION\_REFUSED} $\rightarrow$ \texttt{INCONCLUSIVE} \\
\bottomrule
\end{tabularx}
\end{table}

\section*{Acknowledgements}
AI assistance was used for drafting, language editing, parts of the analysis scripts, and the conceptual illustration in Fig.~\ref{fig:flowpilot_architecture}.

\section*{Code and Data Availability}
The CFD Forge source code, family recipes, validators, prompts, response
schemas, controller-comparison records, figure files, and the scripts that
produce the data figures and tables of this paper are available at\\
\url{https://github.com/aakashoza31/physics-constrained-cfd-agent}.\\
The archived session records, the native case files of every run (including
the precursor channel case that supplies the cube inflow), source snapshots,
solver logs, evidence packets, field-evolution videos, and the provider usage
records are deposited on Zenodo at
\url{https://doi.org/10.5281/zenodo.23148676}.

\end{document}